B. R. Erick Peirson[1,2†] ORCID:0000-0002-0564-9939
Erin Bottino[1]
Julia L. Damerow[1]
Manfred D. Laubichler[1,3,4]

Quantitative Perspectives on Fifty Years of the Journal of the History of Biology

[1]ASU-SFI Center for Biosocial Complex Systems, Arizona State University, Tempe, AZ 85287-4501
[2]arXiv, Cornell University Library, Cornell University, 161 Ho Plaza, Ithaca, NY 14853
[3]Santa Fe Institute, 1399 Hyde Park Rd, Santa Fe, NM 87501
[4]Marine Biological Laboratory, Woods Hole, MA 02543

[†]Corresponding author. brp53@cornell.edu. (360) 531 4474.

**Abstract**. *Journal of the History of Biology* provides a fifty-year long record for examining the evolution of the history of biology as a scholarly discipline. In this paper, we present a new dataset and preliminary quantitative analysis of the thematic content of *JHB* from the perspectives of geography, organisms, and thematic fields. The geographic diversity of authors whose work appears in *JHB* has increased steadily since 1968, but the geographic coverage of the content of *JHB* articles remains strongly lopsided toward the United States, United Kingdom, and western Europe and has diversified much less dramatically over time. The taxonomic diversity of organisms discussed in *JHB* increased steadily between 1968 and the late 1990s but declined in later years, mirroring broader patterns of diversification previously reported in the biomedical research literature. Finally, we used a combination of topic modeling and nonlinear dimensionality reduction techniques to develop a model of multi-article fields within *JHB*. We found evidence for directional changes in the representation of fields on multiple scales. The diversity of *JHB* with regard to the representation of thematic fields has increased overall, with most of that diversification occurring in recent years. Drawing on the dataset generated in the course of this analysis, as well as web services in the emerging digital history and philosophy of science ecosystem, we have developed an interactive web platform for exploring the content of *JHB*, and we provide a brief overview of the platform in this article. As a whole, the data and analyses presented here provide a starting-place for further critical reflection on the evolution of the history of biology over the past half-century.

# 0. Introduction

In a scathing 1990 review, the late historian of science John Farley complained that, "from its first two-issue volume in 1968, through its increase to three issues per year in 1982, until today, *Journal of the History of Biology* has provided an outlet for the self-perpetuating oligarchy of Darwin scholars" (Farley, 1990). "Is this healthy, I wonder?" Farley went on, "Has the profession now reached such a size that the members can afford to speak only to each other?". Farley enumerated a variety of themes and fields that, in his view, had been chronically underserved in the pages of *JHB*, including oceanography, ethology, botany, anatomy, physiology, biochemistry, bacteriology, and others. Worse, Farley seemed to suggest that *JHB* had nearly missed the social turn in the history of science, remaining fixated on "the history of biological concepts."

It is worth considering the most charitable subtext of his assertions: that as the flagship periodical of the field, the contents of *JHB* are a window onto the diversity and the development of the history of biology. Indeed, early reviewers (e.g. Brown, 1968) hailed *JHB* as a signpost for the maturation of the history of biology as a distinct specialization within the history of science. The approaching quinquagenary of that first issue in 1968 is an apt occasion to evaluate some of the trends and tendencies of *JHB* over the past five decades as a way to understand the development of the broader discipline. As historians of science reflecting on our own activities, such retrospectives should both recount and contextualize the development of our field; this will require a conversation between systematic analyses of what we have collectively produced and criticism that puts those productions into their social and historical contexts.

In this paper, we aim to provide a foundation for posing questions about the development of our field by developing a quantitative dataset that describes some aspects of the content of *JHB* over the past fifty years, and analyzing it with the help of various computational methods.[1] We focus on characterizing three dimensions of the journal's content: geographic coverage, taxonomic orientation, and the representation of thematic fields within the history of biology. The primary objective of this project has been feature extraction: using computational tools to count and measure thematically relevant attributes of articles in *JHB*. As a secondary objective, we consider whether there have been clear directional shifts or changes in the diversity of content over time. The results presented here are not intended as a definitive test of specific claims about the history of the history of biology as a discipline—such as the hegemony of the Darwinists, the bandwagon of biochemistry, or the short-shrift given botanical fields—but rather as a starting point to provoke further discussion that incorporates both quantitative and qualitative approaches. In that sense, we used JHB as a "model organism" to demonstrate the possibilities and opportunities of a set of computational and quantitative approaches.

The advent of digital and computational humanities has led many contemporary historians to understand that there might be something worthwhile in these new methods, but many remain skeptical. In this paper we are not settling this debate; rather we demonstrate what can be done with some of these methods in one particular case—the history of a journal over its first half century. But more importantly, we also emphasize that computational approaches always need to be complemented with humanistic practices of interpretation (not unlike the link between bioinformatics approaches and experimentation in the life sciences). To this end we have focused on not only creating our analysis, but also an interactive space for continued experimentation.

In the spirit of open access that drives a lot of work in computational and digital humanities (Laubichler *et al.* 2013) we have made all of our data freely available (Peirson *et al.* 2017a) so that you yourself may experiment and extend and refine the work that we present in this article. Computational approaches to scholarship entail experimentation, incremental improvement, and new standards of transparency throughout the research process. This paper is thus an invitation: to consider some ways in which we can leverage technology to reflect on our own activities and productions, and to participate in developing computer technologies into sophisticated and historically-nuanced methods for data curation and research.

In conjunction with the work presented here, we have developed an interactive online platform to facilitate further exploration of *JHB* over its fifty-year history. *JHB Explorer* (https://jhbexplorer.org) is an open-source web application developed primarily in the Django (Python) web framework, with front-end visualizations developed using the D3 (JavaScript) framework. The application brings together existing bibliographic metadata, the topic and field models described below, as well as a constellation of linked web services in the digital history and philosophy of science ecosystem to provide a platform for content discovery. Source code and documentation for the project is available on GitHub (https://github.com/upconsulting/jhb-explorer) under the GNU General Public License (version 2).[2] At the end of this article we discuss some of the opportunities and challenges for scholarly work of this kind. The application will continue to grow and evolve as new technology and datasets come online. This will include a more comprehensive analysis of history of science fields based on a larger number of publication venues.

---

[1] For an introduction to the history and evolution of *JHB* by current and former editors, see Dietrich (2017) and Allen and Maienschein (2017).

[2] https://github.com/upconsulting/jhb-explorer/blob/master/LICENSE

# 1. Geography

"[*JHB*] is, after all, an American journal." (Farley, 1990; p. 303)

Where in the world has the history of biology taken place? Approaching science as a cultural phenomenon entails a localized view of scientific activity (Ophir and Shapin 1991). A significant outgrowth of the social turn in the history of science has been a so-called "spatial turn" involving contributions from geography and sociology (Finnegan 2008). For example, a special issue of *JHB* published in 2012 explored the relationships between place and scientific practice in North America, with regard to (for example) the geography of research stations (Vetter 2012), conservation of biological significant locales (Alagona 2012, Rumore 2012), and the geographic dimensions of scientific controversy (Bocking 2012). Other work has focused on the collection and transport of biological materials (Evenden 2004, Hung 2016), especially in relation to colonial activity (Schiebinger and Swan 2005). The core intuition of the spatial turn is that cultural contingency entails geographical contingency: science is shaped by the places in which it is practiced and received (Withers 2009), and thus our understanding of science is greatly expanded by considering a broader array of geo-social contexts. As we consider the scope of our own scholarly activity it is therefore valuable to know just what geographical contexts we have considered, and whether or not the spatial turn has indeed precipitated a more geographically diverse perspective within our discipline.

Similarly, the intuitions driving the spatial turn in history of science prompt us to ask about how our own geographies have shaped the scholarship of our field. Where in the world are the narrators of the history of biology? One of us recently participated in a working group tied to the History of Science Society's strategic plan; concerns about the geographic and cultural diversity of our profession were recurring concerns among participants. Participants asked whether HSS is a North American organization, an Anglo- or Euro-American organization, or a global one. Those concerns may arise in equal part from the desire to increase the overall diversity of our professional organizations, as well as from anxieties about the latent occidental and imperial biases of our scholarship. In this section we consider the geography of the *JHB* from two perspectives: what are the geographic locales on which our scholarship has focused, and where in the world are the people who have produced it?

## Geolocating *JHB*

We examined each of the articles in *JHB* over its entire run, and attempted to identify the physical location of the author at the time of publication, and to determine locations that were discussed in the article content. To locate references to locales, we took a single visual pass over each article and noted any references to municipalities, regions, or states, taking care to spend an equitable amount of time on each article. We assume that we found a subset of the total references to locations. We then found the closest match to that location in the GeoNames geographical database (http://www.geonames.org/), and recorded the corresponding Uniform Resource Identifier (URI).[3] The GeoNames web service plays two

[3] For the sake of consistency, we used current location identifiers and geopolitical boundaries, which is in some cases highly anachronistic. For example, the name "Czechia" was only officially adopted by the Czech Republic in 2016, and the Republic itself has only existed since 1993, yet we have used the current term to tag articles published as early as the 1960s that refer to (historical) Czechoslovakia. For the present high-level analysis this does not have a substantial impact on our results or conclusions. In other studies, however, historians incorporating digital geographic data in their research may find it fruitful to use regional databases of historical place names (e.g. The

significant roles in this analysis: it acts as both a place name authority service, providing consistent and unambiguous references to geographic concepts, as well as a data source, providing relations among geographic concepts along with their locations and extensions in space. We aggregated those data to produce a series of visualizations of the geography of the *JHB* in 10-year increments, starting in 1968.

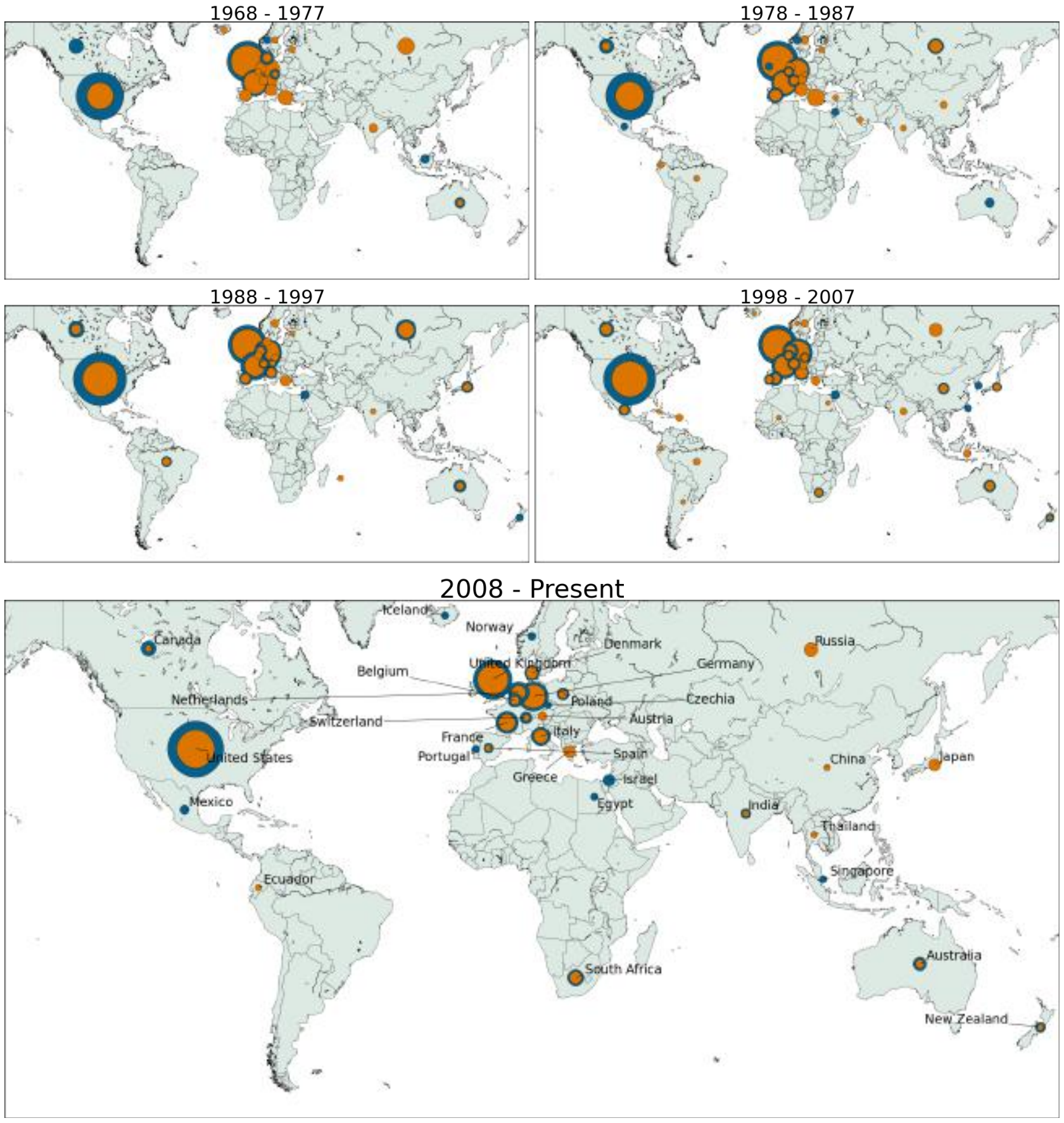

Figure 1. Geographic distribution of *JHB* articles associated by content (orange) and authorship (blue), aggregated by country (using 2016 geopolitical boundaries). Marker radius indicates the relative number of articles associated with each country.

Figures 1 and 2 show the geographic distribution of articles based on content (orange circles) and author (blue halos), aggregated by country. Figure 2 is enlarged to show Europe and surrounding areas. Figure 3 shows the same data expressed as percentages of articles in each period, plotted on a log scale and ordered by descending values for author location. It is important to note that the author location and thematic locations were not readily identifiable for all articles, and so the absence of a location in these visualizations should not be interpreted to mean that no articles were produced in that location nor written about events in that location. Rather, attention should be directed to the relative volume of articles from or about each location, and how those quantities change over time.

Based on our observations, the United States and United Kingdom appear to have dominated *JHB* both thematically and authorially over the entire journal run. In the first decade, the United States was by

---

Historical Gazatteer of England's Place Names [http://placenames.org.uk/]). GeoNames itself also has increasing support for historical place names.

far the largest producer of articles in *JHB*, followed by the United Kingdom (except in the first decade, coming third to Canada). Articles written by historians in Israel appear in the second decade, and become steadily more numerous thereafter. The overall number of countries producing articles increased steadily over the journal run, with nine observed countries in the first decade and twenty-four observed countries in the most recent decade. Over time, we see more articles produced in Asia and the Near-East, and a more even distribution of articles across Europe.

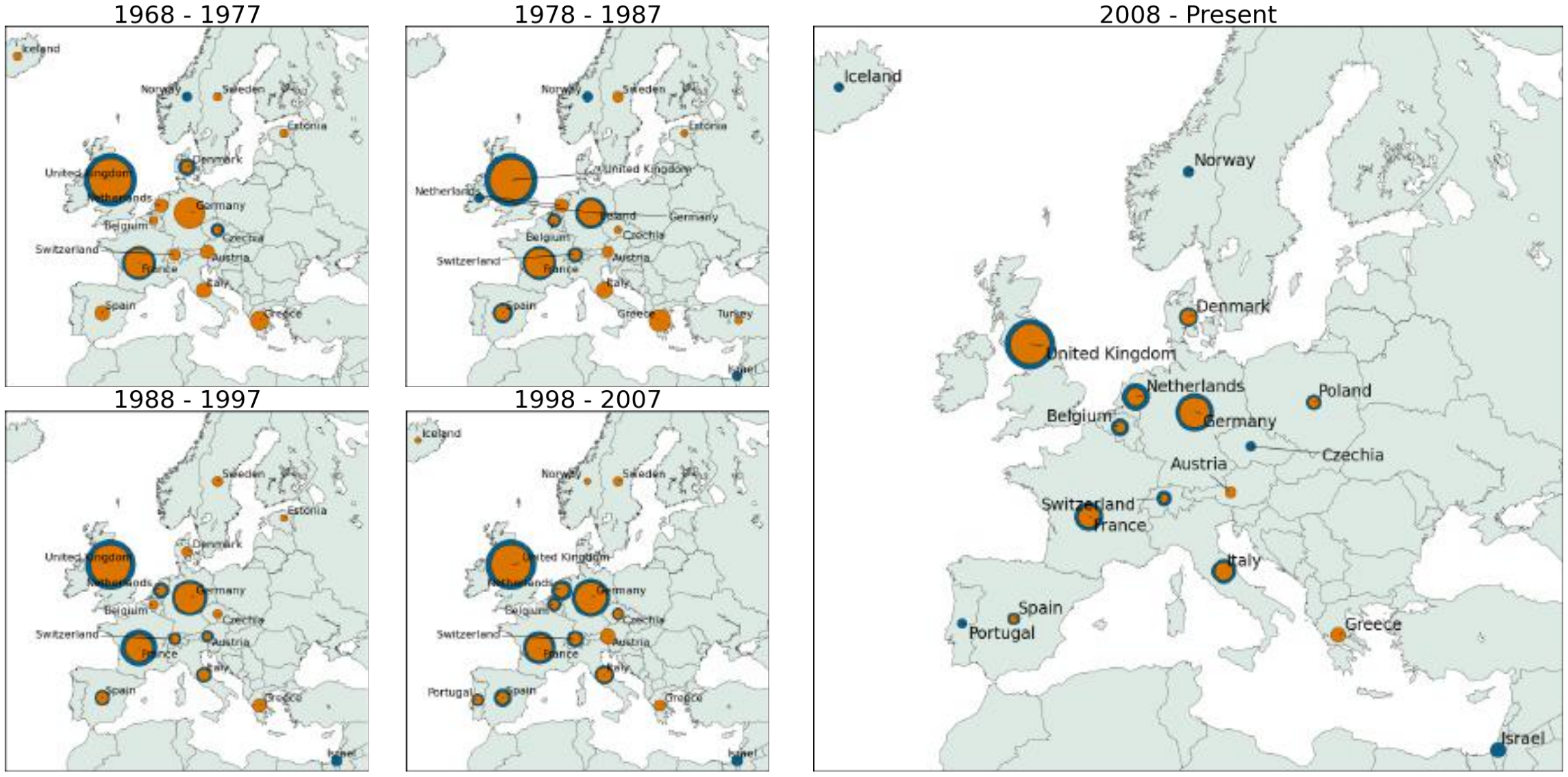

Figure 2. Geographic distribution of *JHB* articles associated by content (orange) and authorship (blue), aggregated by country (using 2016 geopolitical boundaries). Enlarged to show Europe and immediate vicinity. Marker radius indicates the relative number of articles associated with each country.

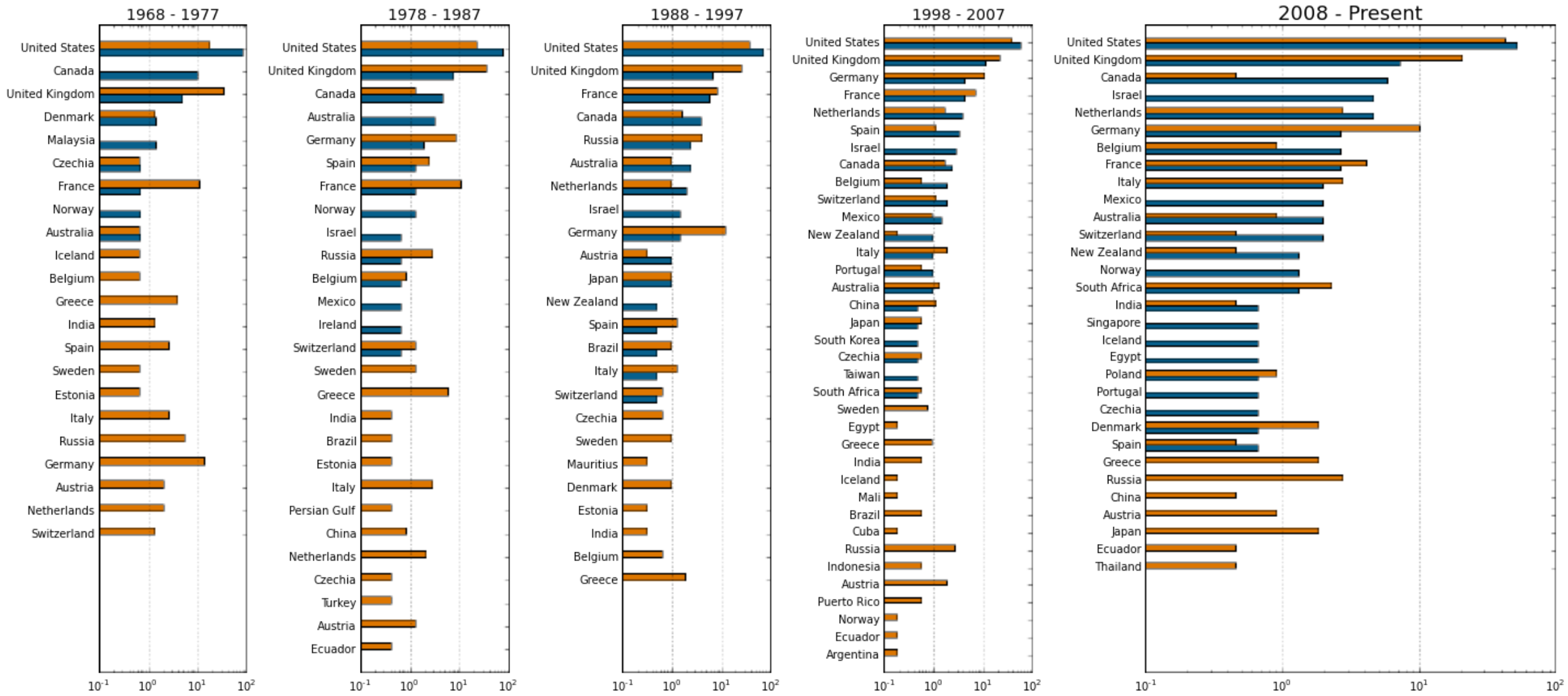

Figure 3. Geographic distribution of *JHB* articles associated by content (orange) and authorship (blue), aggregated by country (using 2016 geopolitical boundaries). Countries are sorted by number articles associated by authorship in each decade. Values are plotted on a log scale.

In the first decade of publication, the United Kingdom dominated *JHB* thematically, followed by the United States, Germany, and France. The United States overtook the United Kingdom in the third decade (1988–1997) and remained the dominant thematic location thereafter. Intriguingly, a high volume of production in a country was not always associated with a high volume of content. For example, Israel was increasingly prominent as a producer of articles over the journal run, but the volume of content concerning Israeli locales was negligible in all decades.

## Geographic Diversity

A significant concern surrounding the geography of our discipline is the geographic diversity of both the content of our research and the practitioners of our trade. In order to make meaningful comparisons over time, it can be useful to translate qualitative conceptualizations of diversity into quantitative metrics. While a comprehensive discussion of the choice and interpretation of diversity metrics from an historiographical context is beyond the scope of this paper, it is important to consider the theoretical motivations for evaluating diversity and how that guides the choice and interpretation of metrics. We therefore apply three different diversity metrics, based on different conceptualizations of what we (as historians of science) might mean by geographic diversity. Each of these metrics greatly oversimplifies the rich body of theoretical work concerning the geography and physical locality of science. Nevertheless, they provide a starting point for thinking about how we can evaluate the geographic scope and distribution of our collective scholarship.

### Shannon Diversity

It is typical to reason about diversity (in general) in terms of both the number of classes represented in a sample—its so-called "richness"—and the evenness of the distribution of things across those samples. We first used the Shannon (1948) index (eq. 1) to evaluate the geographic diversity of *JHB* over time. The Shannon index $H'$ is defined as,

$$H' = -\sum_{i=1}^{R} p_i \ln p_i \qquad \text{(Eq. 1)}$$

where $R$ (richness) is the number of classes, or locations in our case, and $p_i$ is the fraction of instances of the $i^{\text{th}}$ class—articles associated with the $i^{\text{th}}$ location—in the sample. Applied to locations in *JHB* in this way, $H'$ measures the relative uncertainty associated with predicting the location associated with an article drawn at random from a selection of articles. If only one or a handful of locations are present in the sample, then our certainty about our predictions of which location an article will be associated with will be very low; there are, after all, only a few locations to choose from. As the number of locations in a sample increases, and as the distribution of articles across locations becomes more even, our uncertainty about those predictions will increase.

### Simpson's diversity index

Simpson's diversity index is similar to the Shannon diversity index, in that it is based on the relative representation of classes of entities in a sample without regard to relationships among those entities. There are several ways to calculate Simpson's diversity index; in this paper, we use[4]:

$$D_s = 1 - \sum_{i=1}^{R} p_i^2 \qquad \text{(Eq. 2)}$$

[4] Unlike some alternate forms of this index, this form has the convenient property of producing values between 0 and 1.

In the case of geographic diversity, we are once again treating countries as exchangeable; for example, if scientific activity China were discussed in 15% of articles in a given period, it would be numerically equivalent to a scenario in which scientific activity in Brazil were discussed in 15% of articles instead. Unlike Shannon diversity, however, Simpson's diversity index puts more weight on the most prevalent classes of entities—differences in the representation of low-frequency classes will have a smaller effect on the index.

### Geo-proximal diversity

Both Shannon diversity and Simpson's diversity index treat countries as exchangeable classes. In some cases, it is desirable to weight the representation of different classes based on external information, such as their similarity. In the case of geography, we may be interested not only in the representation of various countries, but also how those countries are distributed around the world. For example, we might consider a literature discussing scientific activity in a cluster of seven western European countries to be substantially less diverse than a literature discussing scientific activity in a selection of seven countries spread across the south Pacific, middle east, and north America. We therefore applied a geographic diversity metric adapted from work on Internet topology concerning the geographic diversity of network routes.

Csoma *et al*. (2016) proposed a geographic diversity metric based on the pairwise distances of machines (e.g. switches or routers) along end-to-end routes in the Internet. We apply the low-level component of that metric using distance among geographic points (representations of a geographic locale) within a time period:

$$D_g = \left(1 - \sigma^2_{\Delta'}\right)\bar{\Delta} \qquad \text{(Eq. 3)}$$

where $\Delta$ is the set of pairwise great-circle distances between locales, and $\sigma^2_{\Delta'}$ is the variance of $\Delta$ rescaled to [0, 1]. A collection of points that are far away from each other will have a higher Δ than a collection of points located close together. But if points tend to be clustered together in just a few disparate locations, the mean distances will be discounted by the increased variance in pairwise distances and $D_g$ will be correspondingly lower. We refer to this metric as *geo-proximal diversity*, since we have used the term *geographic diversity* to refer generally to measures of diversity applied to geographic features.

## Summary

The estimated geographic diversity of *JHB* over time is shown in figure 4. Using each of the three metrics described above, we estimated diversity for both the content of articles and for contributing authors. We generated 95% confidence intervals by bootstrap resampling.[5] The geographic diversity of contributing authors shows a clear pattern of steady increase over time under all three metrics. In contrast, changes in geographic diversity in terms of content over time varied substantially among the three metrics. Both the Shannon and Simpson's indices indicated the highest level of diversity in the earliest periods, with a slight dip in 1988–1997 likely due to the relative increase in representation of the United States. Under the Shannon index, there was a slight recovery in the period 1998–2007 probably due to an increase in the number of marginally-represented countries; as we would expect, this was not the case under Simpson's diversity index, which remained lower in the latter periods. On the whole, both metrics indicate that geographic diversity with regard to content was flat or slightly in decline over time, with geographic diversity of authors reaching parity with content only in the most recent period (2008–Present).

[5] Bootstrap resampling is a technique for quantifying uncertainty about an estimated parameter when it is not possible to obtain additional observations. The general procedure involves drawing values at random from a theoretical population modeled on the actual observed sample, on the assumption that the observed sample is "representative" of the population from which it was obtained. This procedure is repeated many times (usually several hundred to several thousand iterations), and the parameter of interest is re-estimated on each iteration. The distribution of parameter values should converge to the theoretical distribution from which the parameter was drawn. We use this technique at several points in this paper.

When we incorporate physical proximity, we see a similar pattern for authorship—lower than content, and increasing steadily over time—but geographic diversity with regard to content is much lower relative to authorship in the earliest periods than it was under the other two metrics. This is probably due to the relatively low representation of Asian and southwest Pacific countries in the earliest periods. Under the geo-proximal metric we see an overall increase in content diversity over time, and authorship diversity reaching parity with content in 1998–2007 or 2008–Present.

Both Shannon diversity and Simpson's diversity index treat countries as exchangeable classes. From an historiographical standpoint, this would be consistent with the view that the geo-social factors acting on science are mostly dependent on the identity of the country in which they take place, rather than economic, political, or social processes acting across national borders. While this may seem an absurd supposition on its face, it may be an appropriate simplifying assumption in cases where regional geo-social processes are less relevant than intranational ones. The geo-proximal diversity index used here, in contrast, heavily weights geographic distance. This may cut with or against the grain of different historiographical frames of reference. For example, proximity will minimize the effect of increased richness due to greater representation of European countries. On the other hand, proximity can be a poor proxy for economic factors (which may be tied to, for example, colonial history); for example, in recent years the economy of South Africa has more in common with the economies of India, Russia, and Brazil (geographically distant) than it does the economy of Mozambique (its immediate neighbor).

If an uninitiated reader were to select a few dozen pages at random from *JHB* in any decade, they could not be faulted for thinking that the history of biology is a story about events and people living in the United States and the United Kingdom and, to a lesser extent, Western Europe. Indeed, judging from the pages of *JHB*, the geographic orientation of the field has remained enormously lopsided. It is unlikely that this geographic bias is due primarily to the geographic distribution of biological research itself. For example, Stocks *et al* (2008) analyzed the geographic distribution of tropical research based on the journals *Biotropica* and *Journal of Tropical Ecology* from 1995 to 2004; based on their data on locale of research (which we would expect *JHB* content to reflect), we would expect to see a Shannon diversity index of 2.82 and a Simpson diversity index of 0.92, both significantly higher than what we observe in *JHB*. Further research based on a much broader survey of biological research is warranted. Yet the geographic richness and diversity of contributing historians themselves is on the rise, and it remains possible that increased representation of non-US/UK locales might bring a more balanced gaze to the field over the next several decades.

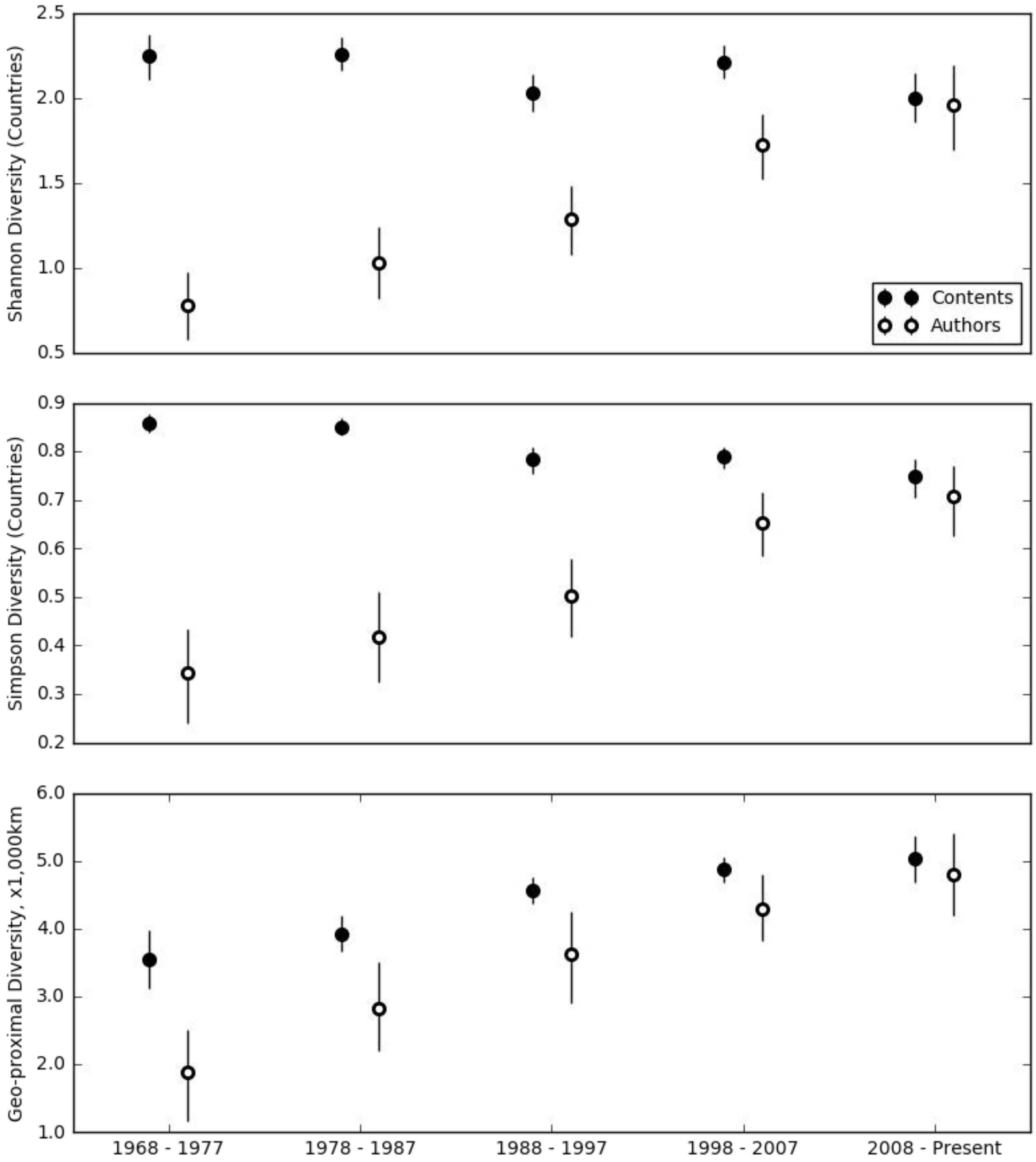

Figure 4. Geographic diversity of the *JHB* over time, in terms of content and authorship, using the Shannon diversity index (eq. 1), Simpson's diversity index (eq. 2), and geo-proximal diversity (eq. 3). From the perspective of the Shannon metric, the geographic diversity of *JHB* in terms of its content remained stable or declined slightly over time, while the geographic diversity of its authors increased steadily and reached parity with content in the most recent period. Simpson's diversity index yields similar results, but with relatively lower diversity in the most recent periods. In contrast, from the perspective of geo-proximal diversity, in which physical proximity is considered, geographic diversity with regard to content did increase over time. The relative significance of these metrics depends on the extent to which one's interpretive frame of reference values physical proximity as a proxy for epistemically-relevant geosocial differences.

# 2. Taxonomy

"Botany too has been ill served..." (Farley, 1990; p.303)

Organisms are a valuable record for the history of the history of biology. Although the extent to which organisms are a focal concern either of the science described or the historian describing that science may vary, it is difficult to conceptualize biological research—and by extension any history of biology—that is completely divorced from organisms. Organisms nucleate research communities (Leoneli and Ankeny 2013), collaborate with scientists to spur new lines of research (Kohler 1994), alter the landscape of plausible explanatory models (Peirson 2015), and mediate interactions among economic, political, and scientific concerns (Rader 2004). The taxonomic orientation of scholarship in the history of biology can thus provide another valuable high-level view onto the attentions of our field.

The availability of robust classification algorithms and public taxonomic databases provide some valuable mechanisms for summarizing the discussion of specific organisms across very large collections of text (Peirson *et al* 2017b). In this section, we describe trends in the taxonomic orientation of *JHB* using methods from Named Entity Recognition (NER). We hope that these data will provide a starting-place for broader reflection on the scope of historical scholarship with respect to biodiversity as well as the status of organisms in the history of biology.

## Named Entity Recognition

We used the LINNAEUS NER model (Gerner *et al* 2010) to tag references to organisms on each page of *JHB*. NER is a problem in information retrieval in which the goal is to identify words or phrases in a text that refer to instances of a particular class of entities, such as people, places, institutions, or dates. NER is usually achieved through supervised machine learning, in which a "training set" of human-annotated documents is used to train a classifier. LINNAEUS is a dictionary-based NER application, in which a large collection of documents from Medline and PubMed Central that had already been tagged with entries from the NCBI Taxonomy database were used to generate a lexicon of phrases that refer to specific taxa. LINNAEUS matches both formal taxonomic terms (e.g. species binomial names) and common names (e.g. "mouse"). Unlike most NER models, which merely focus on individuating entity references, LINNAEUS also disambiguates those entity references against the NCBI Taxonomy database, allowing us to retrieve lineage data for each organism reference.

In total, we found 19,488 references to 867 unique taxa (excluding humans) across the entire journal. 17,904 of those references were identified at the rank of species, and the remaining 1,584 were to subspecies. The distribution of references to individual taxa was extremely long-tailed: 53% of taxa were mentioned only once. We assume that we have found a reasonably representative subset of all of the explicit or implicit references to organisms in *JHB*.

The most frequently mentioned taxon (allowing for multiple mentions on a given page) was *Rattus norvegicus* (996 mentions), followed by *Apis mellifera* (768), *Zea mays* subsp. mays (747), and *Mus musculus* (694). We found it surprising, however, when considering only the number of articles in which a taxon was mentioned that *Equus caballus* (201 articles) was the most prevalent, followed by *Canis lupus familiaris* (186), and *Bos taurus* (130). Table 1 shows the top five organisms mentioned in each of six time periods by both raw reference count and by number of articles. One of the potential pitfalls of dictionary-based NER is that it may produce false-positives when organism names are included in common idioms (“dog tired,” “hobby-horse”), inanimate objects (“saw horse”), or other phrases that are related to the organism but perhaps not a direct reference. As an example of the latter, we were curious to know whether the values for *B. taurus* were artificially inflated by references to bovine serum albumin (BSA), widely used as a protein concentration standard in modern laboratories: in fact, the term "bovine" was found in only 3 of the 554 phrases identified as references to *B. taurus*, suggesting that a significant number of detections are indeed genuine references to the domestic ungulate. In other cases, detected references to the word “horse” were indeed due to idioms. Using the raw detection count may be a more reliable indication of overall (actual) reference volume; when article counts are used, a single idiomatic use of the term “horse” in a two-page book review is given the same weight as a full-length article specifically about the evolution of horses.

Table 1. Top 5 most frequently referenced organisms in each period.

| | References | | | Articles | | |
|---|---|---|---|---|---|---|
| **Period** | **Taxon** | **N** | **%** | **Taxon** | **N** | **%** |
| 1968 - 1975 | *Bos taurus* | 121 | 6% | *Ziziphus mauritiana* | 37 | 30% |
| | *Saccharomyces cerevisiae* | 121 | 6% | *Canis lupus familiaris* | 33 | 27% |
| | *Canis lupus familiaris* | 119 | 6% | *Equus caballus* | 31 | 25% |
| | *Equus caballus* | 110 | 6% | *Ovis aries* | 23 | 19% |
| | *Ziziphus mauritiana* | 92 | 5% | *Bos taurus* | 21 | 17% |
| 1976 - 1983 | *Solanum tuberosum* | 119 | 7% | *Canis lupus familiaris* | 29 | 20% |
| | *Canis lupus familiaris* | 88 | 5% | *Equus caballus* | 27 | 19% |
| | *Equus caballus* | 88 | 5% | *Bos taurus* | 20 | 14% |
| | *Sturnus vulgaris* | 59 | 4% | *Ziziphus mauritiana* | 20 | 14% |
| | *Apis mellifera* | 49 | 3% | *Gallus gallus* | 15 | 10% |
| 1984 - 1991 | *Apis mellifera* | 360 | 16% | *Equus caballus* | 35 | 18% |
| | *Milicia excelsa* | 191 | 8% | *Phleum pratense* | 31 | 16% |
| | *Felis catus* | 166 | 7% | *Canis lupus familiaris* | 25 | 13% |
| | *Zea mays* subsp. *mays* | 80 | 4% | *Alocasia macrorrhizos* | 21 | 11% |
| | *Hemisus marmoratus* | 76 | 3% | *Ziziphus mauritiana* | 18 | 9% |
| 1992 - 1999 | *Mus musculus* | 450 | 11% | *Rattus norvegicus* | 31 | 13% |
| | unidentified phage | 439 | 10% | *Equus caballus* | 27 | 11% |
| | *Rattus norvegicus* | 387 | 9% | *Sus scrofa* | 26 | 11% |
| | *Zea mays* subsp. *mays* | 220 | 5% | *Oryctolagus cuniculus* | 25 | 10% |
| | *Hemisus marmoratus* | 213 | 5% | *Mus musculus* | 25 | 10% |
| 2000 - 2007 | *Rattus norvegicus* | 422 | 10% | *Equus caballus* | 41 | 10% |
| | *Zea mays* subsp. *mays* | 350 | 8% | *Canis lupus familiaris* | 40 | 10% |
| | *Equus caballus* | 266 | 6% | *Bos taurus* | 29 | 7% |
| | *Apis mellifera* | 235 | 5% | *Mus* sp. | 25 | 6% |
| | *Alocasia macrorrhizos* | 151 | 3% | *Oryctolagus cuniculus* | 25 | 6% |
| 2008 - 2016 | *Bombyx mori* | 232 | 5% | *Equus caballus* | 40 | 14% |
| | *Canis lupus familiaris* | 230 | 5% | *Canis lupus familiaris* | 39 | 14% |
| | *Alocasia macrorrhizos* | 221 | 4% | *Gallus gallus* | 32 | 11% |
| | *Bos taurus* | 197 | 4% | *Bos taurus* | 31 | 11% |
| | *Canis rufus* | 138 | 3% | *Felis catus* | 29 | 10% |

## Taxonomic Concentrations

Figure 5 shows the representation of organisms in *JHB* over time in terms of the percentage of articles in each period that refer to a member of a particular phylum. Note that the proportions shown do not sum to 100% within a period, since a single article may mention members of more than one phylum. Between 70% and 83% of articles in each period mentioned a member of Chordata (chordates), between 57% and 75% mentioned a member of Streptophyta (green plants), and between 20% and 30% mentioned a member of Arthropoda (arthropods). Members of Chordata were mentioned more often than any other phylum in all periods except 1984–1991, when Chordata and Streptophyta were evenly represented. All other phyla were mentioned in fewer than 12% of articles in each period.

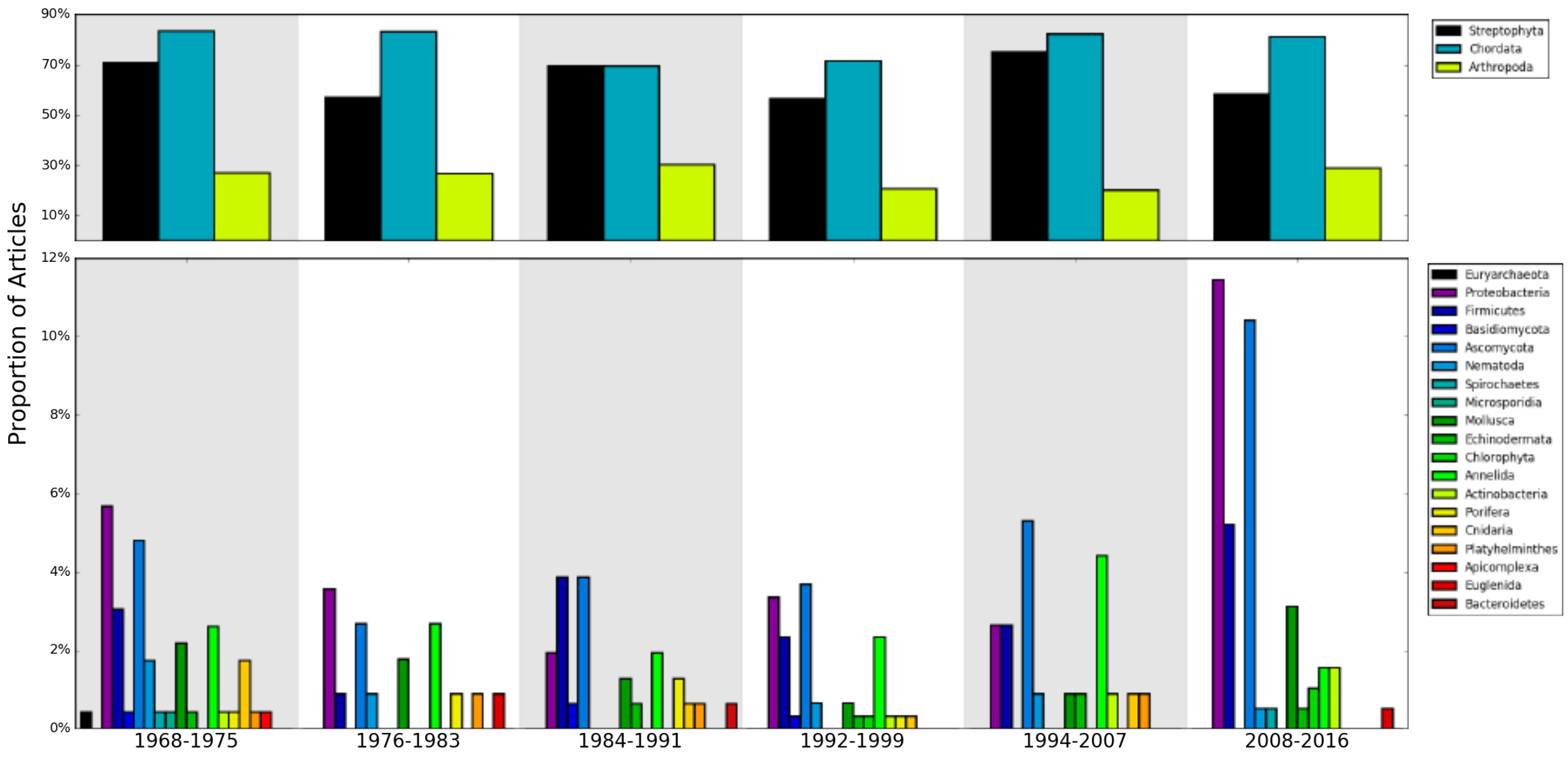

Figure 5. Organisms mentioned in *JHB* over time, grouped by phylum. Organism references were identified using the LINNAEUS Named Entity Recognition model (Gerner *et al* 2010). Values shown are the percentage of articles in a period that refer to a member of each respective phylum. Note that since a single article can refer to members of more than one phylum, the values within each period will sum to more than 100%.

One notable shift in recent years is an increase in the representation of Proteobacteria, from 2–6% of articles in each period prior to 2008 to more than 11% in the period from 2008 to 2016. In that latter period, 18 out of the 22 articles containing references to members of Proteobacteria mentioned

*Escherichia coli*, and 2 mentioned *Lawsonia intracellularis* (which can cause intestinal hyperplasia in pigs). *Pseudoalteromonas atlantica* (which produces marine biofilms), *Serratia marcescens*, and *Haemophilus influenzae* (human pathogens) were each mentioned in a single article. A similar increase can be seen for Ascomycota, rising from 2–6% prior to 2008 to about 10% thereafter. This shift was also driven by a single organism: 19 of the 20 articles mentioning members of Ascomycota in the latter period referred to *Saccharomyces cerevisiae* (brewer's yeast), and 1 mentioned *Aspergillus niger* (a black mold).

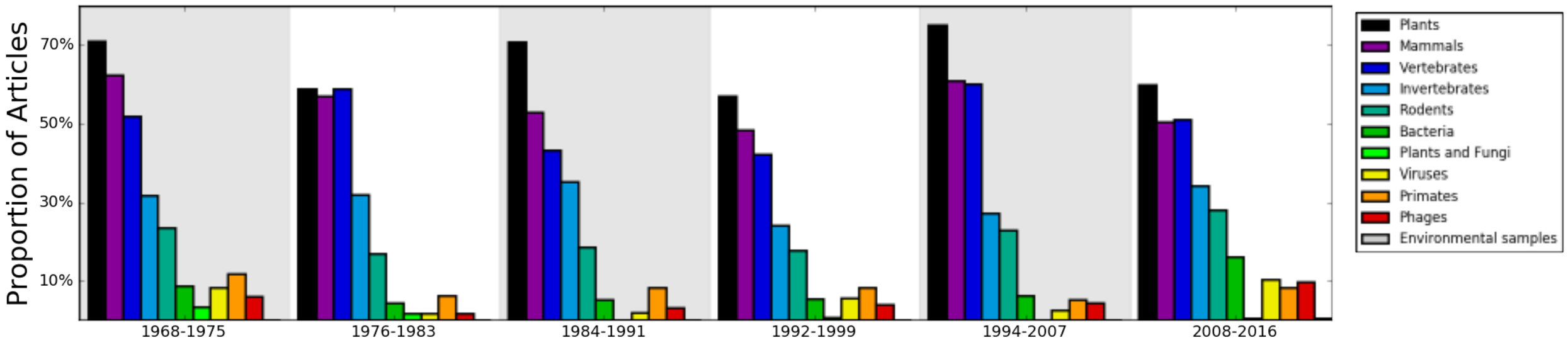

Figure 6. Organisms mentioned in *JHB* over time, grouped by the NCBI Taxonomy database “Division” classification. Although those divisions overlap, each of the taxa identified in *JHB* were assigned to a single division. Therefore, the value shown for Vertebrates is for vertebrate organisms that do not fall into Mammals, Rodents, or Primates; the value shown for Mammals if for mammalian organisms that do not fall into Rodents or Primates, etc.

Figure 6 shows the representation of taxa grouped by the NCBI Taxonomy database's "division" field. Note that although these categories are not mutually exclusive, each taxon reference is assigned to only one division (by design of the Taxonomy database). Thus, the true representation of Mammals is the sum of "Mammals", "Rodents", and "Primates" in figure 6. In *JHB*, the "Plants and Fungi" division is largely represented by *Nicotiana* spp. and *Solanum* spp., but also includes several references to fungi like *Ustilago maydis* (corn smut) and *Aspergillus glaucus*. From this perspective, we again see the relatively high representation of bacteria in recent years compared to earlier periods. Phages are also somewhat enriched in the latter period compared to earlier periods.

## Taxonomic Diversity

We used the taxonomic diversity index proposed by Clarke and Warwick (1998):

$$\Delta_T = \frac{\sum_i \sum_j \omega_{ij} x_i x_j}{n(n-1)/2} \quad \text{(Eq. 4)}$$

where $x_i$ is the abundance of the $i^{th}$ taxon, and $\omega_{ij}$ is the weighted path length between taxon $i$ and taxon $j$ in a taxonomic tree. Clarke and Warwick’s taxonomic diversity index is a particularly appealing tool for diachronic analysis since it is not biased by differences in sample size, allowing for meaningful comparisons among years and subsamples with variable availability. We constructed our tree from lineage data in the NCBI Taxonomy database, excluding non-ranked nodes, using the lowest rank (usually species) for each taxon that we identified in *JHB*. We used bootstrap resampling to account for the potential impact of sample size on diversity in each year.

The taxonomic diversity of organisms mentioned in *JHB* has increased overall since 1968 (Figure 7). That pattern of increase was approximately monotonic up through the period 1992–1999. During the period 2000–2007, however, taxonomic diversity constricted to levels comparable with the period 1976–1983, recovering slightly (but not statistically significantly) in the period 2008–2016.

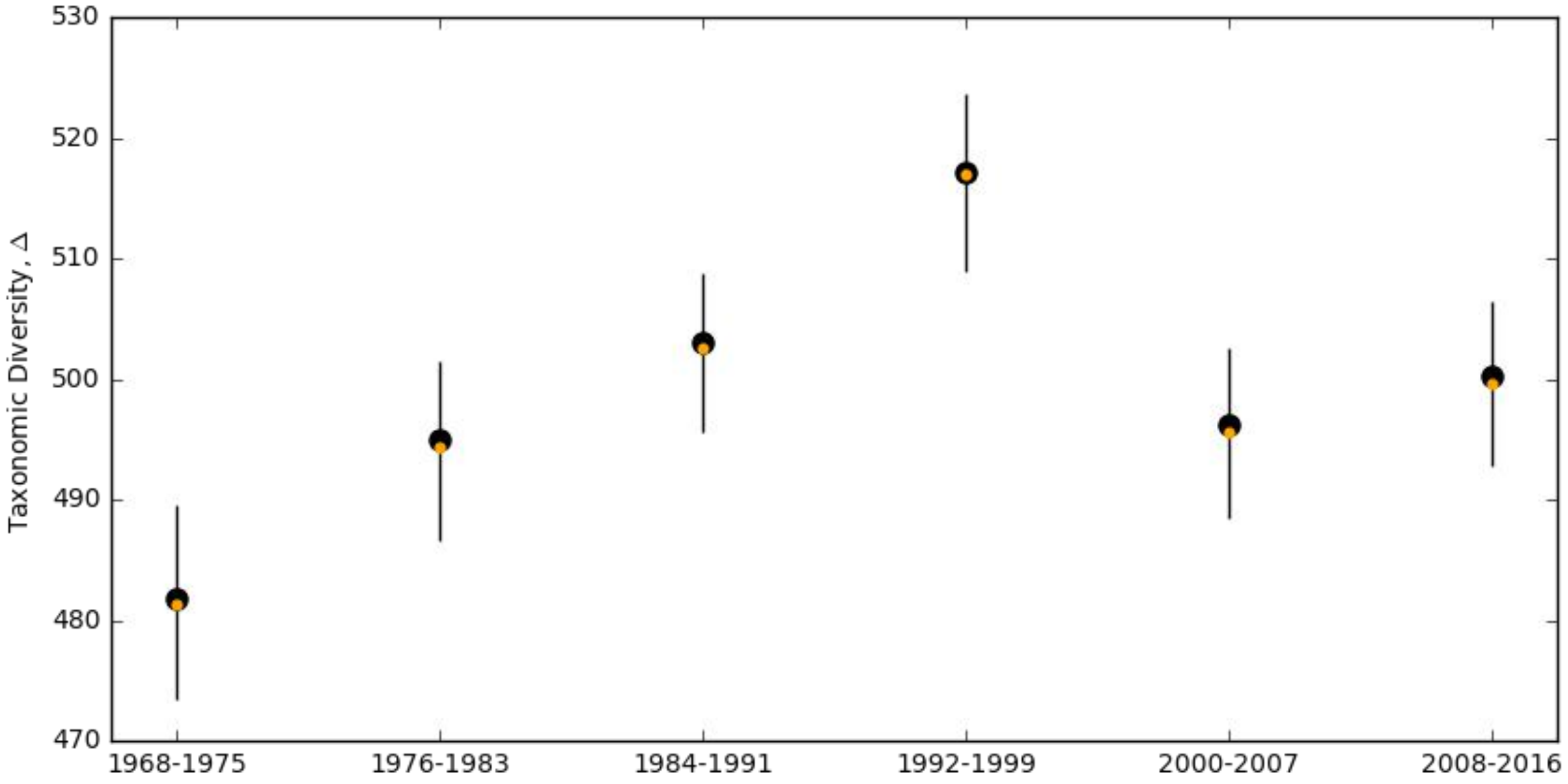

Figure 7. Taxonomic diversity of *JHB* over time, based on mentions of organisms. Diversity was calculated as in Clarke and Warwick (1998; eq. 4), using taxonomic relations from the NCBI Taxonomy database. Error bars show bootstrapped 95% confidence intervals generated by resampling organism references within each period. Taxonomic diversity increased steadily over the first four periods, but declined significantly in the period 2000–2007.

In previous work (Peirson *et al* 2017b) we examined taxonomic diversity in biomedical research literature over the period 1975–2016. We found that diversity had increased steadily from about 450 in the mid-1970s to about 550 in the mid– to late–1990s, and remained relatively stable thereafter. The increase in diversity in *JHB* was somewhat slower than in the broader literature. Taxonomic diversity in *JHB* was higher than in the broader biological research literature in the period 1976–1983 (~494 in *JHB* compared to 450–475 for biomedical literature), but lower in the peak period of 1992–1999 (~515 compared to 450–475). Although results for the biological literature were reported on a more granular temporal scale, it is intriguing to note the correspondence between the post-1999 decline in diversity in *JHB* and the plateau in the broader biological literature in the late 1990s. Further research is needed to explore potential relationships between organism choice in the sciences and what historians of biology choose to write about (and editors choose to publish).

## Summary

Although the historiographical significance afforded to the particular organisms involved in biological research varies widely in both magnitude and kind, biological diversity nevertheless forms an important thread in the meshwork of materials and concepts undergirding the biological sciences. In cases where organisms have been highlighted as participants in scientific activity we have seen evidence that the specific characteristics of organisms can have consequences for the direction of scientific research and the construction of scientific knowledge (Burian 1993). It matters, then, that the histories of biology examine scientific activity across a diverse assemblage of organisms. The evidence presented above suggests that changes in the landscape of biological research during the 1990s—such as the Human Genome Project, and the interest and controversy surrounding model organisms—may have precipitated a more dramatic constriction in the taxonomic diversity of organisms discussed in *JHB* than occurred in the biological literature itself. It is perhaps unsurprising that there would be an interplay between contemporary controversies surrounding biological research and the subject matter on which historians of biology focus (or editors choose to publish). Paying attention to the organismal component of our attentions can help us to effectively criticize and contextualize the scholarly output of our field.

What of Farley's contention, which we have heard repeated by other historians of biology, that botany (or plant sciences more broadly) have been underrepresented in *JHB*? It is important to differentiate between historical literature that primarily focuses on plant-centric disciplines and historical literature in which plant species are mentioned; in this case we have direct access only to the latter. Nevertheless, the results presented here suggest that Streptophytes are consistently the second most frequently mentioned taxon in *JHB*, which seems far from short shrift. Further research should explore in greater detail the connections between taxonomic representation and historiographic shifts within the history of biology.

# 3. Contexts and their Words

> "Social historians concerned with eugenics, birth control, and the growth of institutions, to name but three, are not publishing much in this journal." (Farley, 1990; p. 304)

> "Perhaps one can be pleased that [*JHB*] still is home for those interested in the history of biological concepts." (Farley, 1990; p. 304)

In the previous sections, we described how the content of *JHB* has shifted over time with respect to two specific classes of features: geographic locales and organisms. In this section, we attempt to develop a more general view of the thematic content of *JHB*. We begin by fitting a topic model that provides an abstract description of quasi-independent contexts or thematic elements distributed across individual pages in the journal. That model provides a mechanism for making comparisons among whole articles or between collections of articles (e.g. grouped by publication date, author, or geography) in terms of their thematic content. We then develop a model of higher-level fields by isolating clusters of thematically similar documents across the journal run. We use that field model to illustrate when and how the content of *JHB* has changed over the past five decades.

In the following section, we provide a brief introduction to topic modeling for those who are unfamiliar with the technique. For further discussion of topic modeling in the humanities, including both technical and critical perspectives, see volume 2 issue 1 of *Journal of Digital Humanities* (Meeks and Weingart 2012). For those already comfortable with topic modeling, one can skip ahead to section 3.2.

## 3.1. Topic modeling: an overview

The widespread availability of digitized texts has transformed scholarship in both subtle and dramatic ways. Our ability to index and retrieve relevant content has improved dramatically, expanding the scope of available materials for analysis. At the same time, our ability to work with large collections of texts algorithmically has dangled a carrot in the face of social scientists and humanities scholars: that with the right mathematical tools we might be able to "read" from a distance, and render visible large-scale trends that we might otherwise be too deep in the weeds to recognize. That hope has elevated a family of methods known as "topic modeling"—a branch of information retrieval—to an iconic status within digital humanities. As we discuss at greater length below, it epitomizes a variety of popular ideas about digital culture and "big data" that are potent in the zeitgeist of our time.

The act of finding and bringing together documents that share a particular thematic affinity in a rapidly-spreading morass of literature goes well beyond the history of science. Indeed, online search optimization, specifically the so-called *library problem*, has been one of the defining technical problems of the Google era. The library problem is "to locate ... documents containing information about a particular topic with a fairly high degree of reliability" (Jensen, 1965). Searches based on the presence or absence of words perform relatively poorly on the library problem, as even simple linguistic relations like synonymy and polysemy are not supported. Early internet search indexes like Yahoo addressed this problem through human curation of links in a thematic hierarchy. In many cases, domain-specific indexes like the *Isis Bibliography of the History of Science* (http://data.isiscb.org) are the cutting edge.

Sophisticated unsupervised search algorithms (such as those powering Google) have undergone a flurry of development since the late 1980s. An information retrieval strategy based on the "latent semantic structure" of texts was proposed in 1989 that began to address the problem of synonymy (Deerwester et al, 1989). Latent Semantic Analysis (LSA),[6] as it is now known, nucleated a cluster of probabilistic[7] techniques—commonly referred to as "topic modeling"—that attempt to incorporate something about the "meaning" of words or phrases into document search. The most recent (and well-known) of those techniques is Latent Dirichlet Allocation (LDA), introduced by Blei *et al* (2003). One way to think about topic modeling is as a form of unsupervised machine learning that can be used to find regularities in the distribution of words across a collection of documents. Topic modeling reduces the computational complexity of comparing documents and (ideally) improves the relevance of those comparisons with regard to thematic content of interest to humans.

In the context of digital humanities, one of the main motivations for seeking the kind of corpus-level view of thematic content offered by topic modeling has been to better understand how specific themes or groups of themes are distributed across divisions of the corpus. Those divisions might be temporal—for example, to look for topics that are disproportionately represented at one time or another—or based on some assumptions of the research project, such to compare literature generated in different social or cultural contexts.

The basic strategy of topic modeling is to represent collections of texts in terms of a relatively small number of abstract "topics" that capture the most salient regularities in those texts. If I can transform a search string into such a topical space, then I should be able to find documents that are highly relevant to my query even if none of the specific words in that query occur in those documents. A common way to think about a topic model is as a vector-space model (VSM). A simple vector-space model of a corpus might describe documents as vectors of feature (word) counts; equivalently, each document is represented as a single point in an N-dimensional hyperspace, where N is the number of features in the corpus.[8] A topic model reduces the computational cost of comparing documents by providing a lower-dimensional hyperspace that captures the most salient components of variation in the word-count model.

[6] LSA involves a sparse matrix representation of a corpus in which each cell contains the number of times a word (rows) occurs in a given document (columns; e.g. a page, paragraph, or sentence). A decomposition is applied to the matrix so that the number of rows is reduced while attempting to maintain the similarity relations among columns.

[7] The use of terms like "probabilistic" and "inference" in the topic modeling literature, as well as names like *Latent Dirichlet Allocation* (referring to the Dirichlet distribution), can give the false impression that such methods are designed to facilitate statistical reasoning around knowledge claims. In most cases, those terms are instead a reflection of the fact that underlying computational techniques and algorithms have been borrowed from statistical computing.

Latent Dirichlet Allocation (LDA) is by far the most popular class of topic model in current use. The basic assumptions of LDA are that the contents of a document can be described as a mixture of several topics, and that if a particular topic is present then certain words will be more likely to occur in that document than others. For example, I might reasonably suppose that a document about botany is more likely to contain words like "plant", "meristem", and "sepal", than a document about guinea pig breeding. When we apply this to a specific collection of documents (a corpus), we assume that each word in each document was "generated" by a specific topic. If we can infer which topic generated each word in the corpus, then we can describe our corpus in terms of the representation of topics across those documents.

LDA uses a statistical representation of those assumptions as a computational tool to obtain a description of the topics present in a collection of documents. For each word in each document, we assume that the topic with which the word is associated was drawn from a multinomial distribution, specific to that document, over topics and that the word itself was drawn from a multinomial distribution, specific to that topic, over words. The parameters of those two sets of multinomial distributions (topics given documents, words given topics) are in turn drawn from a Dirichlet distribution—hence the "Dirichlet" in Latent Dirichlet Allocation. Note that in most search applications we don't know ahead of time what topics are present in a corpus, nor what words in particular are most likely to occur in documents that contain those topics. Thus the term *latent*: we believe that some topics (in some sense) exist, but can't actually observe them directly. Instead, we look at how words are distributed across the corpus, and try to obtain the topical structure that best explains that distribution.[9] LDA gives us a model of the document-generating process, and we have some data (texts) produced by that process; our remaining task is to infer the most likely parameters for that model.

One elegant and efficient way to address that kind of problem is by using a Bayesian Markov-Chain Monte Carlo (MCMC) simulation called Gibbs sampling.[10] Bayes' theorem, considered one of the foundational concepts of modern probability theory, tells us that the probability of an hypothesis given some observed data is proportional to the probability of the data given that hypothesis.[11] In a nutshell, Gibbs sampling leverages that relationship to hone in on the most likely hypothesis (usually the value of some parameter, such as a mean) by iteratively proposing new hypotheses and then comparing their relative likelihood given some data. Over successive iterations, the algorithm should "converge" on a distribution of samples that approximates the distribution of credible values for that parameter given the data. MCMC simulations like Gibbs sampling make it possible to work with extremely complex statistical models for which the direct calculation of joint probabilities is intractable.

---

[8] VSMs based on word counts are usually called "bag-of-words" models, since such models do not take into account the order in which words occur within each document.

[9] In most variants of LDA, we decide ahead of time how many topics (usually on the order of tens to hundreds of topics) will be inferred.

[10] A Markov process is one in which the state of a system at a given time depends only on the immediately previous state of the system. "Monte Carlo" refers to the fact that the simulation is stochastic.

[11] We often reason like this in our daily lives. For example, one of us has a dog named Pinyon who appears to be a mix of several breeds: she has the coloration of a red heeler, but is considerably undersized. When we speculate about her pedigree, we ask ourselves: "If a red heeler and corgi were to mate, how likely is it that their offspring would look like Pinyon?" We may not be able to answer that question with a specific probability, but we are reasonably sure that Pinyon is a more likely outcome under the heeler-corgi hypothesis than, say, the hypothesis that her parents were a red heeler and an American bulldog. We prefer the heeler-corgi hypothesis.

Although LDA borrows expressions and tools from Bayesian statistics, it is not a form of statistical data modeling *per se*: it is not designed to be explanatory, nor to evaluate hypotheses. Indeed, it departs from conventional Bayesian data modeling in that the primary concern is not to quantify uncertainty about the model parameters, but rather to achieve a single low-dimensional representation of the corpus. Primarily, LDA is a tool for improving information retrieval, as discussed above. Secondarily, however, the high degree to which the model structure is analogous to everyday intuitions about the thematic content of documents raises the prospect of using those parameters as a high-level quantitative view of a corpus. A topic model is first and foremost a tool for exploration: perhaps to find relevant documents that you didn't already know about, or to provoke questions that can be addressed through further analysis. It can also be used as a dimensionality-reduction technique when asking quantitative questions about the content of a corpus, as we discuss later on.

Since LDA is not focused on quantifying certainty in model parameters, it is acceptable (and popular) to fit LDA models using an algorithm called *collapsed Gibbs sampling* that samples only the topic assignments of the words themselves, with significant gains in computational efficiency. The collapsed Gibbs sampling procedure for LDA involves considering the word-tokens one at a time: For each token, the current assignments of topics to the current document, and the assignments of words across the corpus to the $K$ topics, are used to calculate the relative probabilities that the token was generated by each of the $K$ topics. A new topic is then assigned to the token at random, given those calculated probabilities. Over time, the assignments for individual word-tokens will change less and less frequently, and the model will converge on the most likely state of topic-token assignments. The parameters of interest—the relative representation of topics in each document, and the probability of each word in the vocabulary given a particular topic—can then be calculated from the topics assigned to each word token. The main trade-off of collapsed Gibbs sampling is that we sacrifice the kinds of things that are usually of interest in a statistical data model, like the ability to evaluate hypotheses or make statements about the likely values of model parameters.

After a topic model is fit to a collection of documents, the induced topics are labeled for human inspection by enumerating several of the most probable words in each topic. Those labels are often sufficient to give an impression to the informed reader as to the nature of the topic. In most cases there will be a very large number of words that are highly probable for a given topic, and so one should not assume that the presence of a topic in a document is strictly indicative that those labelling terms are present. For example, a given topic may have several thousand words with non-negligible probability, whereas a page of an article in *JHB*—the unitary "document" in this project—may have fewer than 200 tokens. Thus, even if 95% of the tokens on a page were assigned to a topic labeled, "mendel, gene, genes, breeding, peas," the likelihood that any one of those five specific labeling terms is actually present in the document is relatively low.

## 3.2. Topic Modeling *JHB*

Prior to model fitting we applied several preparatory transformations to the paginated full text provided by JSTOR. Within each document, we removed running headers from each page. We located the likely beginning of the bibliography within each article (if one was present) based on the distribution of specific punctuation characters and key terms; if a bibliography was identified, that content was excluded from analysis. We tokenized each page at the level of individual words, removing all punctuation, whitespace, and numeric characters. Since many concepts of interest in *JHB* are represented by multi-word phrases, we extracted two- to six-word phrases by applying (in three sequential passes) the criterion $\frac{N_{ij}-5}{N_i+N_j} > 0.1N$, where $N_{ij}$ is the number of occurrences of the bigram (word $i$ followed by word $j$), $N_i$ is the total number of occurrences of word $i$, and $N$ is the total number of tokens in the whole corpus (Řehůřek and Sojka 2010). After conjoining word-parts into phrases as described, we removed tokens of any individual words that (a) occur in the Natural Language ToolKit stopwords list (Bird *et al* 2009), or (b) occur on more than 6,000 pages (about 1/4 of the corpus).

We fit a series of topic models to the articles in *JHB*, treating each page as a separate document[12] and varying the number of topics with $K \in \{100,200,300,400,600,800\}$. We used the parallelized collapsed Gibbs sampler implemented in the InPhO Vector Space Model package (Murdock 2015), and ran each simulation for 10,000 iterations. In each case the simulation converged successfully. The discussion below is based on results obtained with $K = 100$. All six models are available for exploration in the *JHB Explorer* platform. The top ten most probable words for each topic in the $K = 100$ model are shown in table 2. While the majority of the topics evoke specific content-related themes, e.g. "linnaeus, classification, plants…" (table 2: 36), we note the presence of several more general topics that may be examples of the "framing topics" described by Priva and Austerweil (2015), e.g. "problem, terms, general, biological…" (table 2: 31).

Our primary objective in topic modeling *JHB* was to gain some insights into how the thematic orientation of the journal may (or may not) have changed over time. Due to the assumption that each page is comprised *entirely* of topics in the model, however, the temporal distribution of individual topics are subject to high multicollinearity. Thus it is not valid to draw inferences directly from an apparent increase or decrease in the representation of a topic over time.[13] We therefore evaluate the distribution of topics across divisions of the corpus in terms of statistical enrichment: that is, how much more or less likely one it is to encounter a particular topic in a given sub-division than one would be expect given the overall probability of encountering the topic in the corpus. Unlike raw topic proportions, enrichment is not subject to high multicollinearity: the enrichment of any one topic in a given division cannot be directly inferred from the enrichment of the other topics in that division.

We divided the corpus into six periods of equal duration, and ranked topics by their relative enrichment—that is, how much more likely a document in that period is to contain the topic than a

[12] Especially in long articles, we assume that topics are not distributed evenly from beginning to end: the author may discuss certain topics early in the article to frame their research, and raise other topics later on. Fitting the topic model at the level of the page better reflects our belief that the probability of a token having being drawn from a particular topic is influenced by the position of the token in the article. The increased corpus size also provides a larger set of "observations" leading to better model fit. We do not lose the ability to consider topic-article assignments, as we can simply combine the token-topic assignments for all of the pages in a document.

[13] In the form of LDA used here, we assume that each page is comprised entirely of topics in the model. In other words, each and every word in the document was generated by one of the K topics. Thus if topic 3 comprises only 32% of a particular document, the remaining 68% of that document must be explained by one or more of the other topics in the model. If topic 95 is less prevalent in documents from 2000 than those from 1999, there will be a corresponding adjustment in the prevalence of the other K-1 topics in that year. This is perfectly reasonable given the objective of providing a more succinct description of the corpus. But it frustrates our ability to draw inferences about processes going on outside of the corpus itself on the basis of the distribution of any given topic. That is not to say that such processes are unrelated, merely that caution should be exercised when interpreting the distribution of topics directly.

Table 2. Top 10 most probable words in each topic, in the 100-topic LDA topic model.

| | |
|---|---|
| 0 | heredity, bateson, inheritance, variation, johannsen, galton, pangenesis, pearson, characters, weldon |
| 1 | human, china, chinese, sollas, humans, human evolution, peking man, origin, modern, skull |
| 2 | kuhn, university, german, institute, unger, berlin, germany, butenandt, professor, laboratory |
| 3 | mayr, ernst mayr, society, synthesis, evolutionary synthesis, genetics, journal, simpson, evolutionary biology, evolutionary |
| 4 | view, concept, views, fact, problem, ideas, idea, although, theories, evidence |
| 5 | dunn, dobzhansky, anderson, goldschmidt, genetics, taxonomy, mayr, gould, experimental taxonomy, dunn papers |
| 6 | dna, molecular biology, protein, rna, watson, crick, proteins, structure, sequence, information |
| 7 | first, scientific, field, darwins, biography, specific, significant, woodger, influence, find |
| 8 | creighton, wheeler, brown, disease, endocrinology, rickets, schafer, creighton papers, ants, physiology |
| 9 | fisher, statistics, goethe, statistical, data, grant, analysis, hensen, variance, pearson |
| 10 | genetics, morgan, gene, genes, chromosomes, drosophila, muller, genetic, heredity, goldschmidt |
| 11 | race, races, knox, racial, human, black, white, anthropology, blacks, hunt |
| 12 | variation, natural selection, selection, change, changes, variations, environment, weismann, process, organisms |
| 13 | forbes, natural history, sprague, expedition, gray, marine, soc, ship, report, british |
| 14 | women, female, male, sex, sexual, woman, behavior, males, females, hormones |
| 15 | eugenics, social, eugenic, heredity, movement, eugenicists, race, davenport, american, society |
| 16 | hooker, social, london, huxley, professional, society, british, natural history, botany, joseph hooker |
| 17 | biochemistry, chemical, biochemical, warburg, chemistry, enzymes, biochemists, metabolism, zymase, mcelroy |
| 18 | laboratory, field, university, thienemann, station, lab, reighard, went, laboratories, place |
| 19 | cuvier, french, paris, lamarck, france, bernard, georges cuvier, plus, general, death |
| 20 | lawrence, anatomy, vivisection, galen, dissection, vesalius, anatomical, medicine, body, medical |
| 21 | insects, kirby, bees, insect, entomology, beaumont, packard, st martin, mantids, hive |
| 22 | animals, macleay, cuvier, animal, blainville, birds, series, owen, fish, hincks |
| 23 | university, students, became, laboratory, department, group, later, teaching, early, physiology |
| 24 | cambridge, zoology, university, bateson, manchester, physiology, london, professor, balfour, ravenel |
| 25 | lorenz, behavior, animals, animal, lorenz s, thorpe, instincts, carpenter, ethology, human |
| 26 | charles, illness, charles darwin, health, family, children, autobiography, his father, death, symptoms |
| 27 | box, committee, support, folder, meeting, society, program, scientists, project, report |
| 28 | chase, roux, south africa, india, land, water, trees, cooper, indian, preservation |
| 29 | volume, historians, chapter, historical, authors, essays, reader, readers, author, chapters |
| 30 | radiation, aec, abcc, radioisotopes, studies, nuclear, atomic energy, neel, effects, genetics |
| 31 | problem, terms, general, biological, way, use, view, point, fact, question |
| 32 | cell, cells, protozoa, cowdry, nucleus, organisms, cell theory, aging, conjugation, growth |
| 33 | aristotle, female, male, semen, parts, animals, generation, material, form, body |
| 34 | experiments, results, used, found, different, studies, could be, using, method, produced |
| 35 | rashevsky, wood, sverdrup, department, evans, vaughan, weaver, stanley, laboratory, folder |
| 36 | linnaeus, classification, plants, genus, genera, system, type, characters, natural, botany |
| 37 | kammerer, russian, institute, psychology, fish, promptov, physiology, przibram, torpedo, moscow |
| 38 | jerne, bordet, ehrlich, serum, antibodies, antibody, ehrlich s, normal, antigen, substances |
| 39 | bronn, fossil record, data, paleontology, fossils, gould, paleontologists, origin, sepkoski, nasa |
| 40 | huxley, owen, haeckel, carpenter, german, spontaneous generation, origin, richard owen, gegenbaur, archetype |
| 41 | simpson, zuckerkandl, goodman, pauling, molecular evolution, molecular, proteins, evolutionary, human, sokal |
| 42 | harvey, fernel, blood, metagons, platypus, language, medicine, fludd, pagel, spiritus |
| 43 | published, papers, journal, paper, articles, publication, works, books, article, number |
| 44 | wallace, voyage, paper, notebook, notes, beagle, bates, henslow, page, written |
| 45 | world, great, people, art, young, white, herschel, family, love, way |
| 46 | social, society, political, human, wells, labor, progress, economic, morris, world |
| 47 | physics, organism, henderson, matter, laws, physical, properties, biological, organization, mechanism |
| 48 | brooks, conklin, oyster, whitman, mbl, american, american biology, morphology, students, loeb |
| 49 | distribution, regions, humboldt, plants, animals, biogeography, region, world, found, climate |
| 50 | vinogradskii, beijerinck, liebig, bacteria, kluyver, pasteur, microbiology, laboratory, soil, chemistry |
| 51 | brain, physiology, nervous system, animal, nerves, physiological, functions, nerve, function, consciousness |
| 52 | figure, images, drawings, illustrations, image, art, visual, tree, diagram, representations |
| 53 | leeuwenhoek, barry, royal society, eggs, observations, animals, spermatozoa, spallanzani, semen, spontaneous generation |
| 54 | animals, animal, rats, aquarium, king, blyth, mice, birds, mellen, little |
| 55 | agassiz, girard, briggs, baird, fishes, louis agassiz, monsters, jordan, nuclear transplantation, schultz |
| 56 | wright, hogben, tansley, clements, vegetation, canguilhem, fisher, survey, environment, concept |
| 57 | spemann, needham, ltp, learning, memory, organizer, induction, waddington, hamburger, harrison |
| 58 | heart, blood, galen, harvey, arteries, veins, body, liver, aristotle, circulation |
| 59 | plants, plant, leaves, holmes, de bary, leaf, potato, roots, soil, schimper |
| 60 | kellogg, blakeslee, toyama, van beneden, civet, silkworm, bionomics, silkworms, silk, loeb |
| 61 | wolff, haller, buffon, descartes, generation, maupertuis, matter, bonnet, particles, newton |
| 62 | breeding, breeders, corn, farmers, mexico, varieties, agricultural, inbreeding, breed, agriculture |
| 63 | selection, population, dobzhansky, muller, populations, wright, genes, genetic, fitness, natural selection |
| 64 | natural history, museum, specimens, naturalists, collections, birds, collection, museums, alexander, american |
| 65 | osborn, gregory, cope, nat, amer, paleontology, fossil, gould, hist, soc |
| 66 | molecular biology, protein synthesis, monod, biochemistry, synthesis, monod s, rna, french, proteins, biochemists |
| 67 | german, freud, germany, fischer, jewish, racial, nazi, poll, institute, political |
| 68 | animals, individuals, varieties, animal, plants, different, certain, forms, produced, form |
| 69 | different, fact, point, form, way, general, case, since, certain, possible |
| 70 | letter, letters, wrote, may, june, hooker, march, published, april, january |
| 71 | immunology, burnet, self, cells, immune system, cell, antibodies, antigen, jerne, immunological |
| 72 | virus, viruses, phage, bacteriophage, delbruck, bacteria, dulbecco, genes, beadle, bacterial |
| 73 | oka, soviet, military, sverdlovsk, russian, state, anthrax, smithsonian, soviet union, weiner |
| 74 | weismann, morgan, regeneration, wilson, cell, chromosomes, nucleus, egg, cells, heredity |
| 75 | way, fact, yet, still, perhaps, make, point, without, great, indeed |
| 76 | mendel, mendel s, de vries, hybrids, experiments, plants, characters, hybrid, traits, white |
| 77 | mitchell, boyer, williams, mitochondria, membrane, cell, palade, atp, oxidative phosphorylation, mechanism |
| 78 | cancer, disease, medical, rous, clinical, diseases, medicine, human, studies, alcohol |
| 79 | embryology, form, haeckel, morphology, ontogeny, phylogeny, morphological, embryo, phylogenetic, adult |
| 80 | romanes, metchnikoff, wyman, dalcq, aging, holliday, cells, metschnikoff, orgel, llr |
| 81 | knowledge, use, practices, laboratory, scientists, different, within, particular, field, techniques |
| 82 | university, journal, department, this paper, thank, netherlands, printed, kluwer academic publishers, philosophy, papers |
| 83 | coon, religion, religious, putnam, kingsley, hutton, education, textbook, darwinism, mivart |
| 84 | kihara, japan, milstein, genetics, biophysics, plant breeding, japanese, technology, technique, cell |
| 85 | temminck, birds, gould, specimens, galapagos, islands, thoroddsen, island, finches, beagle |
| 86 | historical, historians, social, scientists, role, within, studies, field, early, period |
| 87 | buffon, azara, daubenton, spain, spanish, natural history, histoire naturelle, lacepède, birds, french |
| 88 | fish, salmon, conservation, fisheries, environmental, management, wildlife, scientists, wolves, wilderness |
| 89 | natural selection, darwinism, spencer, origin, progress, darwinian, evolutionary, evolutionary theory, ruse, darwinian revolution |
| 90 | de vries, moll, czermak, mutation theory, physiological, hugo de vries, went, pangenes, du boisreymond, crustacea |
| 91 | goethe, german, form, individual, kant, whole, beurlen, individuals, generations, forms |
| 92 | god, world, design, creation, laws, natural theology, mind, human, religion, natural |
| 93 | lyell, lamarck, geology, earth, vestiges, chambers, geological, principles, creation, evidence |
| 94 | ecology, ecological, ecologists, ecosystem, adams, odum, ecosystems, community, studies, hutchinson |
| 95 | origin, natural selection, charles darwin, malthus, divergence, principle, notebooks, essay, variation, read |
| 96 | genetics, haldane, lysenko, soviet, party, lysenko s, soviet union, political, darlington, lysenkoism |
| 97 | theophrastus, dioscorides, ex herbis, plant, plants, pliny, greek, ancient, used, tree |
| 98 | blood, oxygen, water, hemoglobin, henderson, barcroft, physiological, physiology, haldane, diffusion |
| 99 | wilson, sociobiology, hamburger, altruism, bacteria, group, price, hadley, culture, kin selection |

Table 3. Top 5 most enriched topics for each period, in the 100-topic LDA topic model.

| | |
|---|---|
| 1968–1975 | blood, oxygen, water, hemoglobin, henderson, barcroft, physiological, haldane, physiology, diffusio |
| | biochemistry, chemical, biochemical, warburg, chemistry, enzymes, biochemists, metabolism, zymase, mcelro |
| | wolff, haller, buffon, descartes, generation, maupertuis, matter, bonnet, particles, newto |
| | creighton, wheeler, brown, disease, endocrinology, rickets, schafer, creighton papers, ants, physiolog |
| | leeuwenhoek, barry, royal society, eggs, observations, animals, spermatozoa, spallanzani, semen, spontaneous generatio |
| 1976–1983 | theophrastus, dioscorides, ex herbis, plant, plants, pliny, greek, ancient, tree, use |
| | temminck, birds, gould, specimens, galapagos, islands, thoroddsen, island, finches, beagl |
| | origin, natural selection, charles darwin, malthus, divergence, principle, notebooks, essay, variation, rea |
| | lyell, lamarck, geology, earth, vestiges, chambers, geological, principles, creation, evidenc |
| | leeuwenhoek, barry, royal society, eggs, observations, animals, spermatozoa, spallanzani, semen, spontaneous generatio |
| 1984–1991 | insects, kirby, bees, insect, entomology, beaumont, packard, st martin, mantids, hiv |
| | harvey, fernel, blood, metagons, platypus, language, medicine, fludd, pagel, spiritu |
| | charles, illness, charles darwin, health, family, children, autobiography, his father, death, symptom |
| | ecology, ecological, ecologists, ecosystem, adams, odum, ecosystems, studies, community, hutchinso |
| | selection, population, dobzhansky, muller, populations, wright, genes, genetic, fitness, natural selectio |
| 1992–1999 | molecular biology, protein synthesis, monod, biochemistry, synthesis, monod s, rna, french, proteins, biochemist |
| | jerne, bordet, ehrlich, serum, antibodies, antibody, ehrlich s, normal, antigen, substance |
| | virus, viruses, phage, bacteriophage, delbruck, bacteria, dulbecco, genes, beadle, bacteria |
| | immunology, burnet, self, cells, immune system, cell, antibodies, antigen, jerne, immunologica |
| | cancer, disease, medical, rous, clinical, diseases, medicine, human, studies, alcoho |
| 2000–2007 | breeding, breeders, corn, farmers, mexico, varieties, agricultural, inbreeding, breed, agricultur |
| | de vries, moll, czermak, mutation theory, physiological, hugo de vries, went, du boisreymond, crustacea, pangene |
| | cambridge, zoology, university, bateson, manchester, physiology, london, professor, balfour, britis |
| | radiation, aec, abcc, radioisotopes, studies, nuclear, atomic energy, neel, effects, genetic |
| | german, freud, germany, fischer, jewish, racial, nazi, poll, institute, politica |
| 2008–2016 | first, scientific, field, darwins, biography, specific, woodger, significant, influence, fin |
| | bronn, fossil record, data, paleontology, fossils, gould, paleontologists, origin, sepkoski, nas |
| | genetics, haldane, lysenko, soviet, lysenko s, party, soviet union, political, darlington, lysenkois |
| | human, china, chinese, sollas, humans, human evolution, peking man, origin, modern, skul |
| | chase, roux, south africa, india, land, water, trees, indian, cooper, preservatio |

document chosen at random from any period—in that subset of the corpus. Table 3 shows the top five most "characteristic" topics for each period.

## 3.3. A cluster-based field model

The LDA topic model discussed in the previous section provides an atomistic description of the content of *JHB* in terms of quasi-independent thematic contexts or topics. After fitting the model, we find that we can retrieve fairly coherent and meaningful topics: *letter*, *letters*, *wrote*, *may*, and *june* (table 2: 70) do indeed seem like words that would be likely to occur when discussing correspondence; *natural selection*, *darwinism*, *spencer*, *origin*, *process* (table 2: 89) do indeed seem like words that would be likely to occur when discussing Darwinian evolution; etc. As we described in section 3.1, the LDA topic model describes each document as a mixture of those thematic kernels. Topics are by hypothesis independent: from the perspective of the model, part of what makes each document unique is that they combine different topics in various proportions.[14] In this section, we use our topic model of *JHB* to develop a higher-level model that individuates distinct discourses or fields characterized by specific combinations of topics. In the same way that individual documents can be described as a mixture of topics, we assume that documents belonging to a specific field will be more likely to contain some topics than other topics. For example, a discourse surrounding the scientific legitimation of racist eugenics in Nazi Germany might be more likely to contain topics concerning eugenics, public policy, and Germany, but less likely to contain topics concerning botany, protozoa, or fisheries.

Earlier, we described topic models as an hyperspace in which each document is represented as a single point. A result of this perspective is that we can use simple vector-based distance measures to compare documents in terms of their thematic content. If the model is effective at providing a meaningful description of the thematic content of the corpus, then documents with similar content will be located relatively close together in that hyperspace. Given our assumption about the relationship between fields and topics, if there are higher-order regularities in the corpus such as discourses or fields that can be described in terms of specific combinations of topics, we should expect to find local structures in the larger topical hyperspace that correspond to those fields.

We identified clusters of thematically similar documents in two steps: first, we found an optimal embedding of documents from the topic model hyperspace into a two-dimensional Euclidean plane; second, we used K-means clustering to identify groups of documents located closely together in the two-dimensional embedding.

### 3.3.1. t-Distributed Stochastic Neighbor Embedding

Stochastic Neighbor Embedding (Hinton and Roweis 2002) is one of a family of algorithms for non-linear dimensionality reduction (NLDR). Like their linear counterparts (e.g. Principal Component Analysis), NLDA methods are designed find low-dimensional embeddings of high-dimensional data that preserve the implicit structure of a dataset (de Silva and Tenenbaum 2003). Popular NLDR algorithms include Isomap (described by Balasubramanian and Schwartz 2002) and Laplacian Eigenmaps (Belkin and Niyogi 2003). The basic approach of SNE is to represent distances between individual data points in both the original high-dimensional space and a low-dimensional space in probabilistic terms, and then attempt to find a transformation that minimizes the Kullback-Leibler divergence (Kullback and Leibler 1951)[15] between those two projections of the data (Hinton and Roweis, 2002). t-Distributed SNE (t-SNE) is a variant of SNE that uses a different probabilistic representation of distances between data points—student's t-distribution instead of Gaussian distribution, hence the name—and a more computationally favorable "cost" function (the derivatives of which are used to find the optimal transformation) (van der Maaten and Hinton 2008). We chose to use t-SNE over other NLDR techniques because of its ability to preserve structure on multiple scales. Indeed, we expect high-level thematic structures to have a multi-scale structure: some documents will be more thematically similar to each other than others, and some clusters of similar documents (possibly representing fields) will be more similar to each other than others.

[14] For example, it seems reasonable to us (you should go and check) that our model would describe the essay review titled "Sociobiology: twenty-five years later" by Michael Yudell and Rob DeSalle (2000) as being mostly about sociobiology (29%) and historiography (20%), with some attention to eugenics (8%), and a sprinkling of other topics oriented toward theories, concepts, and argumentation.

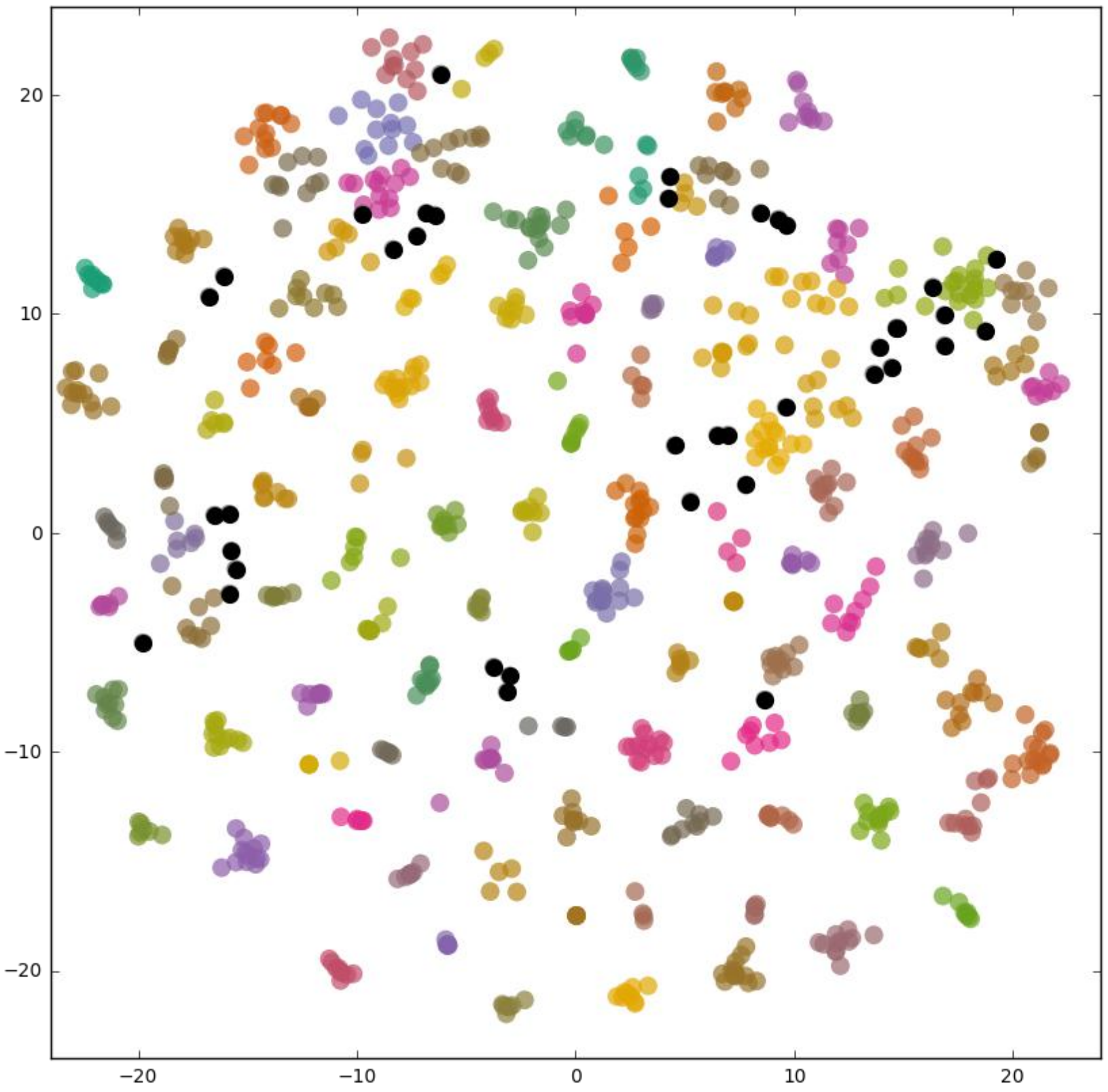

Figure 8. Thematic article clusters within *JHB* (full-length articles and essay reviews). We fit a LDA topic model with 100 topics to all articles in *JHB*, treating individual pages as documents. After model fitting we aggregated topic-page assignments into topic-document assignments, and generated a two-dimensional projection of the corpus using t-distributed Stochastic Neighbor Embedding using cosine distances between articles in the topic model hyperspace. We then optimized a k-means clustering model in the t-SNE space to individuate cohesive thematic fields. Shown here are individual articles (excluding single-book reviews) projected in the t-SNE embedding space, colored by field assignment. Articles shaded black were not assigned to any field.

We used the implementation of t-SNE in Pedregosa *et al* (2011), which uses the Barnes and Hut (1986) algorithm, to find a two-dimensional embedding of the topic representations from the LDA topic model for full-length articles and essay reviews. We used cosine (angular) distance to measure the pairwise differences among documents in the topic model hyperspace. Since the t-SNE algorithm is initialized randomly, we ran the algorithm using a range of seeds and chose the transformation with the lowest KL-divergence (the optimal representation of the structure of the corpus in topical hyperspace). Each time, we ran the algorithm for 1,000 iterations of gradient descent with a learning rate of 1E3, stopping after 30 iterations of no change in the projection state. The optimal projection (KL divergence of 0.84) is shown in figure 8. The most similar documents are most tightly grouped in the 2-D projection and, although less tightly constrained, dissimilar documents tend to be much further apart.

[15] K-L divergence measures the disagreement between two probability distributions.

### 3.3.2. K-Means clustering for thematic fields

Once we obtained an optimal two-dimensional projection of the corpus, we used K-Means clustering to individual groups of thematically similar documents. K-means clustering attempts to segregate data in a hyperspace by assigning each point the one of K clusters with the nearest mean. The algorithm begins by proposing an initial partitioning of the data, obtaining the mean of the data in each partition, and then updating the assignment for each point to the partition with the nearest mean.[16] Note that the number of clusters must be specified ahead of time. Choosing too few clusters may fail to detect meaningful partitions of the data; choosing too many clusters may lead to overfitting—e.g.in the case that K is equal to the number of documents, the optimal result is the one in which each cluster is comprised of a single document. We used a variation on the Akaike Information Criterion (AIC) (Akaike 1973) as a heuristic to guide our selection of K. Since we can think of the cluster partitions/means as a model of the data in its 2d projection, we interpret the distance between each data point and the centroid the cluster to which it is assigned as a residual of the model. Thus we calculate the residual sum of squares for a model with $k$ clusters as:

$$RSS_k = \frac{1}{N_d} \sum_{c \in C} \sum_{d \in D_c} \| \bar{D}_c - d \|^2 \quad \text{(Eq. 5)}$$

where $C_k$ is the set of $k$ clusters and $D_c$ is the set of documents (represented as points) in cluster $c$. Since we want $k$ such that $RSS_k$ and $k$ are simultaneously minimized, we choose $k' = \arg\min_k \ln RSS_k + 2 \ln k$. Since K-means clustering is initialized with a random segmentation of the data, it is possible to arrive at slightly different cluster assignments across different runs of the algorithm, with corresponding variation in $RSS_k$ for a given $k$. Consequently, we performed five rounds of clustering for each $k$, and used the mean value of $RSS_k$ in the objective minimization to obtain the optimal number of clusters. In order to ensure robust cluster assignments, we considered a pair of documents to be co-clustered only if they were both assigned to the same cluster across at least 15 of 20 further runs of the K-means algorithm using the optimal $k$ selected in the previous step. Of those "robust" clusters, we further disregarded any clusters smaller than four documents.

The final field model consists of 104 document clusters, plus a 105th group of non-clustered documents. We considered the possibility that our procedure had simply retrieved the original structure of the topic model, with each cluster merely corresponding to a single prominent topic. We calculated the probability that cluster assignments could be correctly predicted from the most prominent topics in each document. Predicting field assignments based on the single most prominent topic in each document had a success rate of only 68%, indicating that the field model does indeed capture a distinct component of variation in the distribution of content across the corpus. We estimate the overall fit of the field model by interpreting the cosine distance between a document in the topic model hyperspace and the mean vector of its assigned cluster as a residual of the model:

$$dist(\theta_u, \theta_v) = \frac{\theta_u \theta_v^{\mathsf{T}}}{\|\theta_u\| \|\theta_v\|} \quad \text{(Eq. 6)}$$

$$r^2 = 1 - \frac{RSS}{TSS} = \frac{\sum_{c \in C} \sum_{d \in D_c} dist(\theta_d, \overline{\theta}_c)^2}{\sum_{d \in D_c} dist(\theta_d, \overline{\theta})^2} = 0.91 \quad \text{(Eq. 7)}$$

where $D_c$ are the documents in cluster *c*. In other words, the field model explains about 91% of the variance in the thematic orientation of documents in the original topic model hyperspace.

[16] We used the spectral clustering procedure implemented in the SciKit-Learn Python package (Pedregosa *et al* 2011), which uses Lloyd's (1982) algorithm for K-means clustering. Lloyd's algorithm differs from the original K-means algorithm in that each partition is represented as a Voronoi cell encapsulating its assigned data points; whereas the original K-means procedure calculates the mean of each partition based on the data, Lloyd's algorithm uses the centroid of the Voronoi cell.

### 3.3.3. Representation of thematic fields in *JHB*

After fitting the field model as described above, some shifts in the representation of individual fields were readily apparent. Since the articles in *JHB* are only a subset of the total production by historians of biology at any given time, we considered the possibility that the apparent temporal asymmetries that we observed in the distribution of individual fields are due to chance alone rather than a reflection of directional shifts in the orientation of *JHB* (e.g. due to editorial intent, or broader shifts in the discipline). We quantified temporal bias in terms of the relative density of documents in a given field that occur on or after year *t*.:

$$\delta_f(t) = \sum_{t'=t}^{t_{max}} \frac{N_{ft'}}{N_{t'}} - \sum_{t'=t_0}^{t-1} \frac{N_{ft'}}{N_{t'}} \quad \text{(Eq. 8)}$$

where $N_{f,t}$ is the number of documents in field *f* and year *t*, and $N_t$ is the total number of documents (across all fields) published in year *t*. Thus $\delta_f(t) = 1$ and $\delta_f(t) = -1$ indicate that all documents in field *f* occur on or after *t* and prior to t, respectively, and indicates that the documents in field *f* are distributed evenly about *t*. For a field that is distributed in a precisely uniform manner over time, $\delta_f(t) = -\frac{2(t-t_0)}{t_{max}-t_0} + 1$; similarly, if fields are distributed normally about the median year $\frac{t_{max}-t_0}{2}$, then the mean across all fields, $\overline{\delta}(t)$, should have the same shape. The variance in $\delta$ across all fields for a given year corresponds to the overall (a)symmetry of the distributions of the whole collection of fields over time. We estimated the expected variance in $\delta$ by calculating $\delta_f(t)$ using 1,000 randomly generated permutations of $\frac{N_{ft'}}{N_{t'}}$ within each field over time.[17]

Figure 9 shows the observed and theoretical variance and mean for over time. The observed variance in $\delta_f(t)$ falls outside the expected range under the non-directional hypothesis in a broad period around the midpoint of the journal run. As a complementary approach, we also estimated the "half-life" (which we denote as $t_{1/2}$) for each field (i.e. the year in which $\delta_f(t) = 0$) for both the observed and theoretical (permuted) data; variance in $t_{1/2}$ across all fields fell well outside the expected values under the non-directional hypothesis. Both of those results suggest that the relative representation of fields in the journal over time were driven either by changes in editorial decision-making or by directional changes in the overall orientation of the discipline. In other words, the journal moved away from publishing articles in some fields and toward publishing articles in other fields, and did so in a non-random manner.

[17] A permutation test of this kind requires us to assume that the relative proportion of documents among fields as a whole (i.e. across all years) is acceptably representative of the true distribution of scholarly effort among those fields within the discipline.

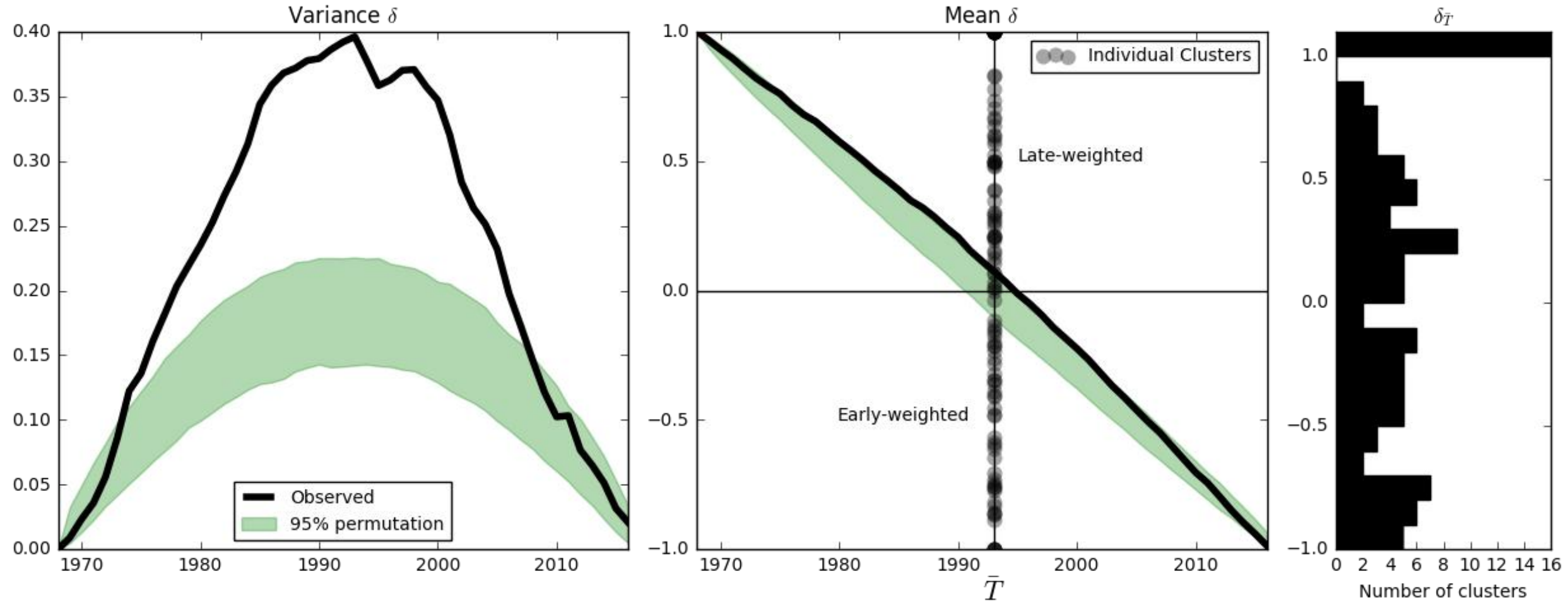

Figure 9. Temporal bias in field representation over time. $\delta_f(t)$ measures the distribution of articles in a given field $f$ occurring prior to and after time $t$ (eq. 8). In a given year, articles from fields with low $\delta_f$ will occur mostly after that year, and articles from fields with high $\delta_f$ will occur mostly prior to that year. The left panel shows variance across clusters in observed $\delta_f$ over time (black line) and the variance that we would expect if the observed temporal distribution of articles within each field were due to chance alone (shaded green area). The middle panel shows the observed and expected mean $\delta_f$ across clusters over time. In the middle panel, observed values of $\delta_f$ at $\bar{T} = 1993$ (the midpoint of the time series, and also the point at which variance in observed $\delta_f$ is maximally divergent from our expectations) are overlaid as dark gray circles, with clusters biased toward later time periods failing in the upper half and those biased toward earlier time periods falling in the lower half. A histogram of observed $\delta_f$ in 1993 is shown in the right panel. These data show that the distribution of individual thematic clusters over time is likely non-random: in other words, changes in journal content are consistent with what we would expect to emerge from scholarly fashions rather than simply expanding the depth or breadth of research in the field.

Table 4 shows fields that are temporally biased, grouped by the period in which each field's respective $t_{1/2}$ falls. Each field is shown overlaid on the two-dimensional projection shown in figure 8. The distribution of each cluster over time (as a proportion of documents in that year) is shown to the lower-right. For each cluster, we calculated the probability of the observed topic assignments for documents in that cluster given the overall distribution of topics in the corpus. On the right-hand side of each cluster's panel are shown the top 5 least probable (i.e. statistically enriched) topics for the cluster. In order to better understand the overall thematic orientation of each cluster, we extracted the top eighteen keywords for each cluster by ranking words based on their $\chi^2$ statistic, which calculated for in-cluster vs. out-of-cluster documents, and the proportion of documents in the cluster in which each word appears. Key words are shown on the left-hand side of each cluster's panel. Whereas we can interpret the enriched topics (right) as responsible for distinguishing their respective cluster of documents from the rest of the corpus, the extracted keywords (left) indicate the shared idiosyncrasies of the documents within each cluster.

Table 4. Thematic fields with temporally biased distributions, grouped by period.

## 1968–1975

leeuwenhoek; spermatozoa; spallanzani; animalcules; royal society; swammerdam; eggs; semen; worms; adb; spontaneous generation; hartsoeker; preformation; spermatozoon; de graaf; egg; leeuwen hoek; testicles;

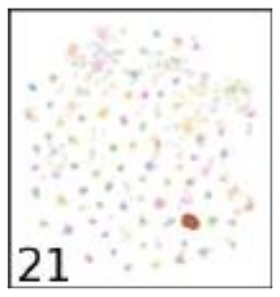

0.00: leeuwenhoek, barry, royal society, eggs, observations
0.00: wolff, haller, buffon, descartes, generation
0.00: historical, historians, social, scientists, role
0.00: box, committee, support, folder, meeting
0.00: aristotle, female, male, semen, parts magendie; flourens; sherrington; spencer; brain; nerve; reflex; gall; cerebrum; nervous system; cerebellum; contraction; psychology; sensibility; muscle; jackson; spinal; consciousness; goodsir;

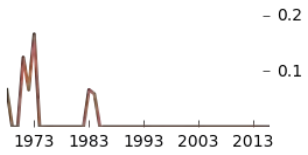

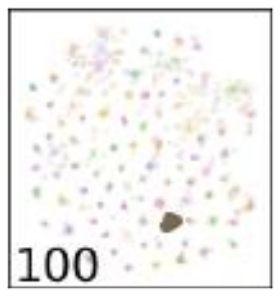

0.00: brain, physiology, nervous system, animal, nerves
0.00: heart, blood, galen, harvey, arteries
0.00: cuvier, french, paris, lamarck, france
0.01: wolff, haller, buffon, descartes, generation
0.01: box, committee, support, folder, meeting wolff; haller; descartes; bonnet; buffon; newton; newtonian; maupertuis; particles; soul; leibniz; boerhaave; hailer; cheyne; generation; epigenesis; adelmann; roger;

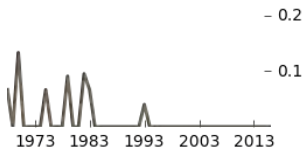

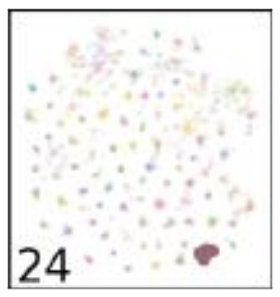

0.00: wolff, haller, buffon, descartes, generation
0.00: leeuwenhoek, barry, royal society, eggs, observations
0.00: god, world, design, creation, laws
0.00: heart, blood, galen, harvey, arteries
0.00: historical, historians, social, scientists, role zymase; buchner; buchner s; yeast; chem; fermentation; warburg; ges; ber; e buchner; hans buchner; protoplasm; ferments; respiration; deut; enzyme theory; fiske; meyerhof;

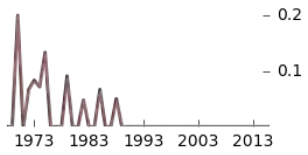

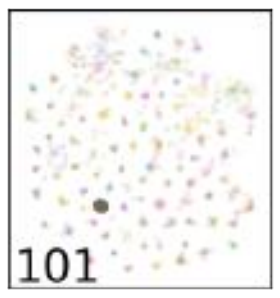

0.00: biochemistry, chemical, biochemical, warburg, chemistry
0.00: vinogradskii, beijerinck, liebig, bacteria, kluyver
0.00: historical, historians, social, scientists, role
0.01: jerne, bordet, ehrlich, serum, antibodies
0.02: mitchell, boyer, williams, mitochondria, membrane lawrence; wallace; breeds; races; breed; varieties; blyth; offspring; prichard; page; peculiarity; domesticated animals; cattle; william lawrence; natural selection; draft; lvo;

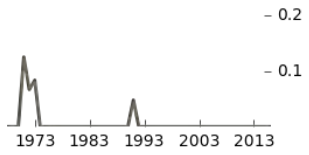

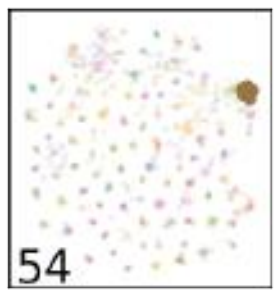

0.00: wallace, voyage, paper, notebook, notes
0.00: origin, natural selection, charles darwin, malthus, divergence
0.00: historical, historians, social, scientists, role
0.00: heredity, bateson, inheritance, variation, johannsen
0.00: mendel, mendel s, de vries, hybrids, experiments

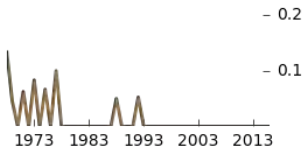

## 1976–1983

sprague; erlanger; gasser; isaac sprague; audubon; pls; plates; journals; citations; asa; charlesworth; magazine; weissenborn; gray; nd session; rd congress; papers; hilde proescholdt; joseph erlanger;

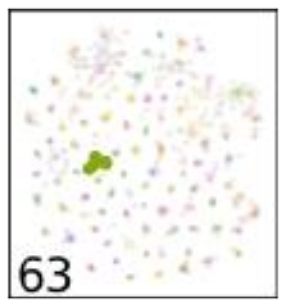

0.00: spemann, needham, ltp, learning, memory
0.00: view, concept, views, fact, problem
0.01: problem, terms, general, biological, way
0.01: wallace, voyage, paper, notebook, notes
0.01: historical, historians, social, scientists, role magendie; flourens; sherrington; spencer; brain; nerve; reflex; gall; cerebrum; nervous system; cerebellum; contraction; psychology; sensibility; muscle; jackson; spinal; consciousness; goodsir;

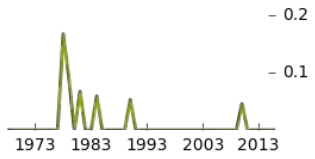

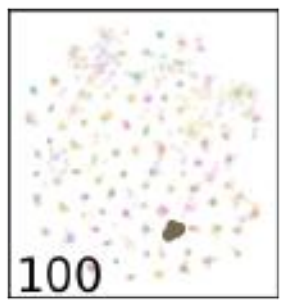

0.00: brain, physiology, nervous system, animal, nerves
0.00: heart, blood, galen, harvey, arteries
0.00: cuvier, french, paris, lamarck, france
0.01: wolff, haller, buffon, descartes, generation
0.01: box, committee, support, folder, meeting theophrastus; dioscorides; ex herbis; poppy; opium; pliny; herbs; opium poppy; herbal; herbals; mandrake; fruit; ex herbis femininis; nicander; datepalm; male; assyrian; hort;

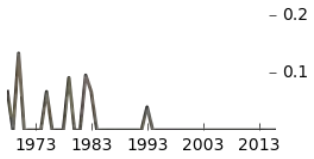

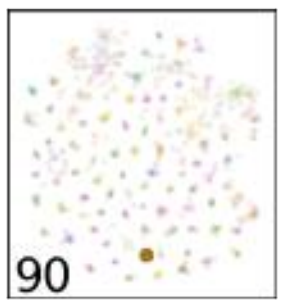

0.00: theophrastus, dioscorides, ex herbis, plant, plants
0.00: historical, historians, social, scientists, role
0.01: figure, images, drawings, illustrations, image
0.02: university, students, became, laboratory, department
0.05: aristotle, female, male, semen, parts

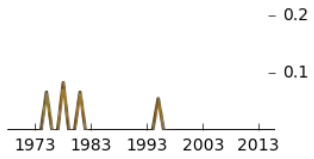

## 1984–1991

ecology; ecologists; ecological; odum; allee; warming; ecosystem; ecologist; animal ecology; evolutionary ecology; emerson; ecosystems; population ecology; volterra; ecological research; adams; nicholson; ibp;

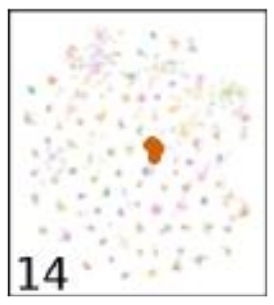

0.00: ecology, ecological, ecologists, ecosystem, adams
0.00: selection, population, dobzhansky, muller, populations
0.01: letter, letters, wrote, may, june
0.01: forbes, natural history, sprague, expedition, gray
0.01: wright, hogben, tansley, clements, vegetation bajema; ghiselin; carl bajema; isis cumulative; barzun; printout; hypotheticodeductive method; yvette conry; j hist; scientisthistorians; jonathan hodge; sir gavin de beer; mankind; ospovat; herovillain; predestinarianism; ebenstein; kavaloski; kivy;

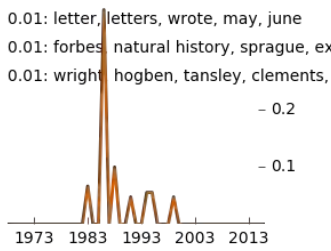

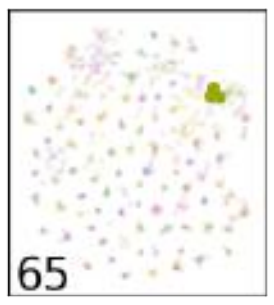

0.00: origin, natural selection, charles darwin, malthus, divergence
0.00: race, races, knox, racial, human
0.00: natural selection, darwinism, spencer, origin, progress
0.04: volume, historians, chapter, historical, authors
0.06: box, committee, support, folder, meeting

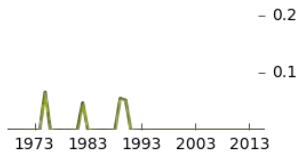

## 1992–1999

photosynthesis; evans; rashevsky; weaver; department; birth control; sverdrup; photosynthetic; wood; vaughan; stanley; scripps; hartman; revelle; series d; at berkeley; spectroscopy; folder;

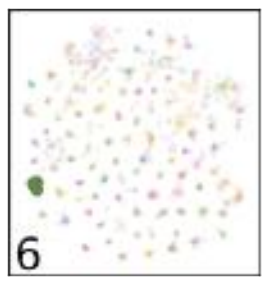

0.00: rashevsky, wood, sverdrup, department, evans
0.00: vinogradskii, beijerinck, liebig, bacteria, kluyver
0.00: molecular biology, protein synthesis, monod, biochemistry, synthesis
0.00: biochemistry, chemical, biochemical, warburg, chemistry
0.00: university, journal, department, this paper, thank sex; women; femininity; hormones; female; male; sex hormones; woman; behavior; androgens; masculinity; sexes; sexual behavior; males; brain; lashley; sex differences; female sex hormones; masculine;

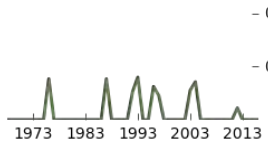

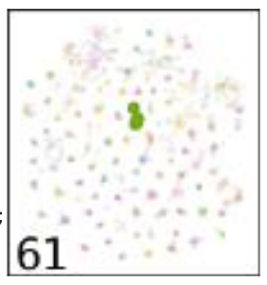

0.00: women, female, male, sex, sexual
0.00: eugenics, social, eugenic, heredity, movement
0.01: social, society, political, human, wells
0.02: romanes, metchnikoff, wyman, dalcq, aging
0.02: race, races, knox, racial, human

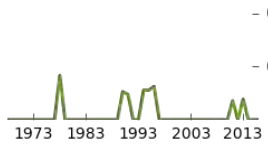

oka; soviet; sverdlovsk; zapovedniki; treaty; russian; soviet union; anthrax; nature protection; meselson; weiner; semenovtianshanskii; gorki; yellow rain; minzoku; bwc; ussr; nature preserves;

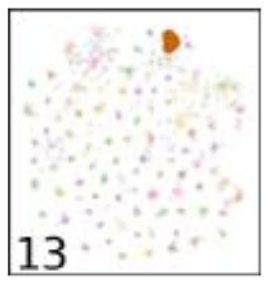

0.00: oka, soviet, military, sverdlovsk, russian
0.00: fish, salmon, conservation, fisheries, environmental
0.00: chase, roux, south africa, india, land
0.00: social, society, political, human, wells
0.00: kammerer, russian, institute, psychology, fish huxley; hooker; letter; vestiges; lyell; darwinism; romanes; gray; vogt; lubbock; chambers; tyndall; mivart; ravenel; bastian; `` historical sketch; letters;

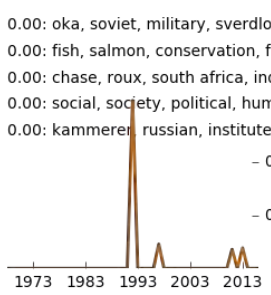

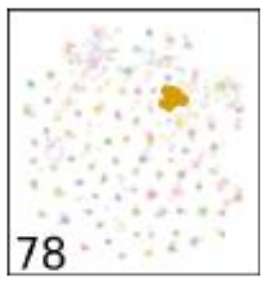

0.00: hooker, social, london, huxley, professional
0.00: natural selection, darwinism, spencer, origin, progress
0.00: huxley, owen, haeckel, carpenter, german
0.00: letter, letters, wrote, may, june
0.00: wallace, voyage, paper, notebook, notes virus; bacteriophage; viruses; delbruck; phage; bacteria; viral; dulbecco; bacterial; bacterium; lysogeny; phages; luria; bacterial viruses; bacteriophages; temin; prophage; dna; dherelle;

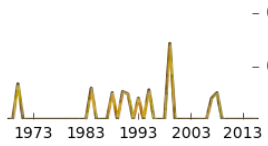

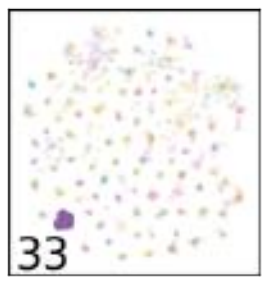

0.00: virus, viruses, phage, bacteriophage, delbruck
0.00: genetics, morgan, gene, genes, chromosomes
0.00: molecular biology, protein synthesis, monod, biochemistry, synthesis
0.00: cancer, disease, medical, rous, clinical
0.01: jerne, bordet, ehrlich, serum, antibodies monod s; monod; protein synthesis; zamecnik; amino acids; rna; incorporation; molecular biology; messenger rna; gros; lwoff; pasteur institute; gale; transfer rna; dgrst; hoagland; paul c zamecnik; soluble rna;

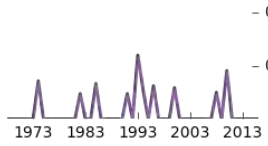

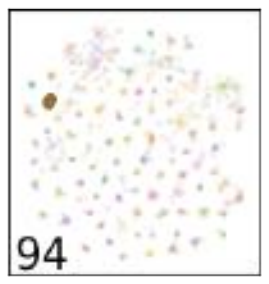

0.00: molecular biology, protein synthesis, monod, biochemistry, synthesis
0.00: dna, molecular biology, protein, rna, watson
0.00: mitchell, boyer, williams, mitochondria, membrane
0.01: biochemistry, chemical, biochemical, warburg, chemistry
0.01: knowledge, use, practices, laboratory, scientists

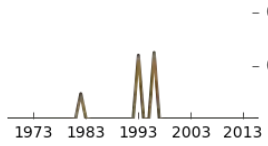

## 2000–2007

scriver; shermer; raby; nance; dolly; cloning; clow; becker; wilmut; mr everyman; specious present; ivorytower syndrome; drake; genetic screening; zinder; everyman; kolata; taboo;

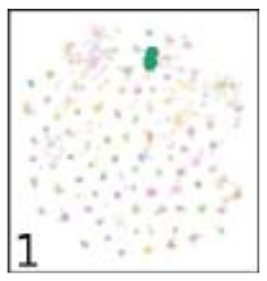

0.00: view, concept, views, fact, problem
0.01: charles, illness, charles darwin, health, family
0.01: first, scientific, field, darwins, biography
0.01: creighton, wheeler, brown, disease, endocrinology
0.02: volume, historians, chapter, historical, authors kuhn; butenandt; schiemann; unger; schramm; virus research; ephestia; euratom; embo; gottingen; institute; alfred kuhn; dahlem; deichmann; wettstein; bonnevie; buzzatitraverso; university;

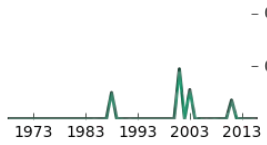

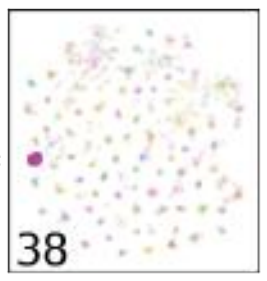

0.00: kuhn, university, german, institute, unger
0.02: box, committee, support, folder, meeting
0.02: goethe, german, form, individual, kant
0.03: rashevsky, wood, sverdrup, department, evans
0.03: mitchell, boyer, williams, mitochondria, membrane hooker; joseph hooker; william hooker; professional; miall; amateurs; kew; professionalism; amateur; drayton; yorkshire; xclub; huxley; jdh; professionals; desmond; xclubbers; gentlemen;

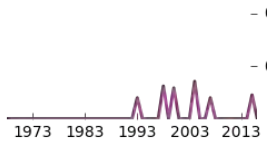

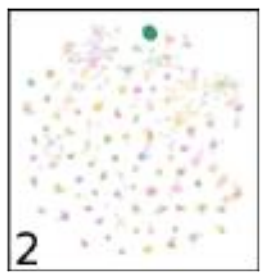

0.00: hooker, social, london, huxley, professional
0.01: knowledge, use, practices, laboratory, scientists
0.01: experiments, results, used, found, different
0.01: world, great, people, art, young
0.01: natural history, museum, specimens, naturalists, collections czermak; lecture hall; anschauung; microphotography; du boisreymond; de vries; lachez; moll; disk; seat rows; frog heart; pangenes; intracellular pangenesis; purkinje; spectatorium; leipzig; kremer; microphotographs; wundt;

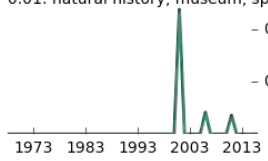

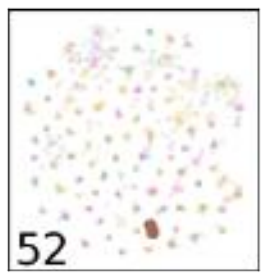

0.00: de vries, moll, czermak, mutation theory, physiological
0.00: goethe, german, form, individual, kant
0.01: figure, images, drawings, illustrations, image
0.04: historical, historians, social, scientists, role
0.05: different, fact, point, form, way

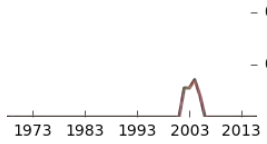

sociobiology; wilson; kin selection; wilsons; sociobiology study group; sociobiological; finnish; ssg; sociobiologists; hadley; lohnis; sociobiology debate; altruism; human sociobiology; finland; multilevel selection; inclusive fitness; selfish;

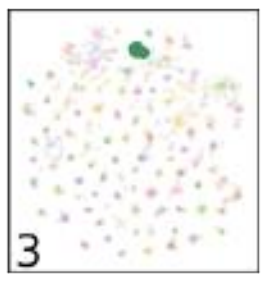

0.00: wilson, sociobiology, hamburger, altruism, bacteria
0.00: social, society, political, human, wells
0.01: lorenz, behavior, animals, animal, lorenz s
0.01: genetics, haldane, lysenko, soviet, lysenko s
0.01: cell, cells, protozoa, cowdry, nucleus aec; radiation; abcc; radioisotopes; atomic energy; neel; isotopes; oak ridge; hiroshima; nagasaki; nas archives; schull; hollaender; exposure; atomic; nuclear; atomic energy commission; manhattan project; iodine;

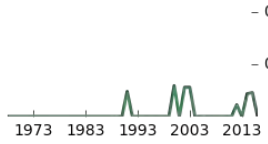

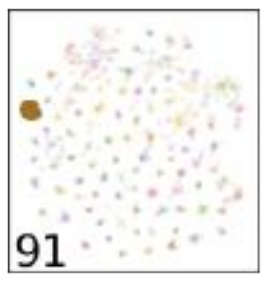

0.00: radiation, aec, abcc, radioisotopes, studies
0.00: rashevsky, wood, sverdrup, department, evans
0.00: molecular biology, protein synthesis, monod, biochemistry, synthesis
0.00: knowledge, use, practices, laboratory, scientists
0.00: oka, soviet, military, sverdlovsk, russian burnet; immune system; immunology; self; antigen; antibodies; lymphocytes; antibody; thymus; immunological; tolerance; immune; jerne; immune response; cst; repertoire; cells; fenner;

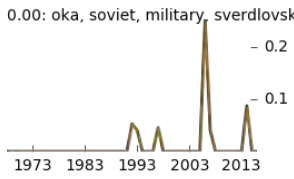

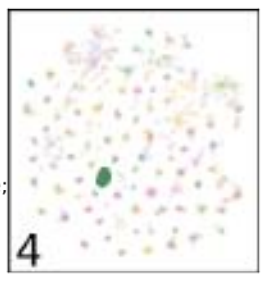

0.00: immunology, burnet, self, cells, immune system
0.00: jerne, bordet, ehrlich, serum, antibodies
0.00: cancer, disease, medical, rous, clinical
0.01: knowledge, use, practices, laboratory, scientists
0.02: kihara, japan, milstein, genetics, biophysics plant breeding; kihara; blakeslee; genetics; biophysics; wheat; japanese; japan; agricultural; plant breeders; vilmorin; ustc; technology; de vilmorin; agricultural practice; rimpau; facs; afbaps; fldr;

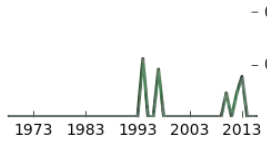

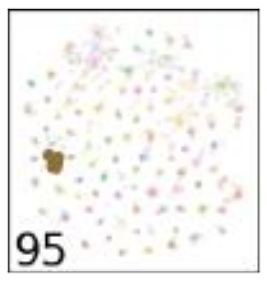

0.00: kihara, japan, milstein, genetics, biophysics
0.00: kellogg, blakeslee, toyama, van beneden, civet
0.00: breeding, breeders, corn, farmers, mexico
0.00: mendel, mendel s, de vries, hybrids, experiments
0.01: genetics, morgan, gene, genes, chromosomes breeders; breeding; inbreeding; farmers; bakewell; colling; breed; horses; corn; crossing; shorthorn; orel; hybrid corn; pedigrees; thoroughbred; bulls; breeding stock; progeny testing; sebright;

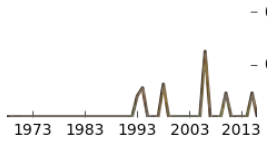

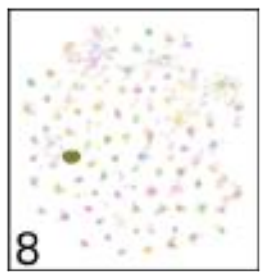

0.00: breeding, breeders, corn, farmers, mexico
0.00: animals, individuals, varieties, animal, plants
0.02: mendel, mendel s, de vries, hybrids, experiments
0.05: university, journal, department, this paper, thank
0.06: hooker, social, london, huxley, professional secord; darwinian revolution; vestiges; lamarck; gautieri; bowler; darwinism; historians; greene; desmond; chambers; evolutionism; bowlers; revolution; victorian sensation; vucinich; `` evorevo; readers; lynch;

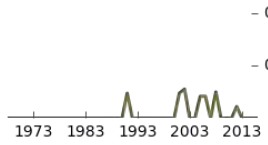

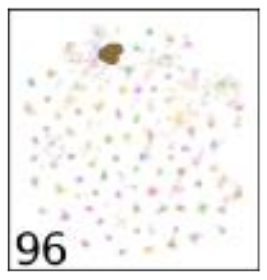

0.00: natural selection, darwinism, spencer, origin, progress
0.00: lyell, lamarck, geology, earth, vestiges
0.00: volume, historians, chapter, historical, authors
0.00: huxley, owen, haeckel, carpenter, german
0.00: coon, religion, religious, putnam, kingsley

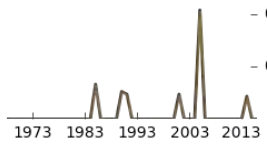

2008–Present

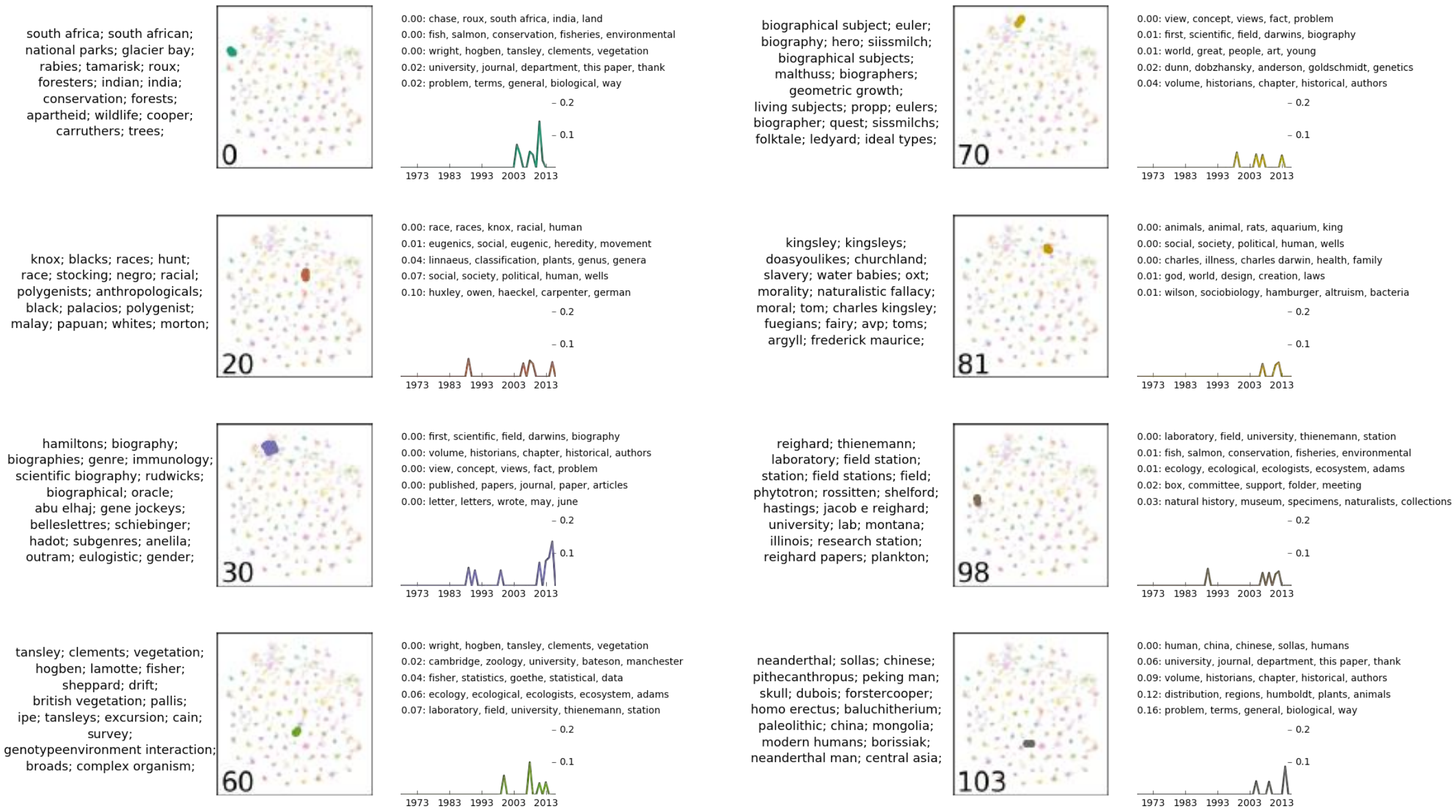

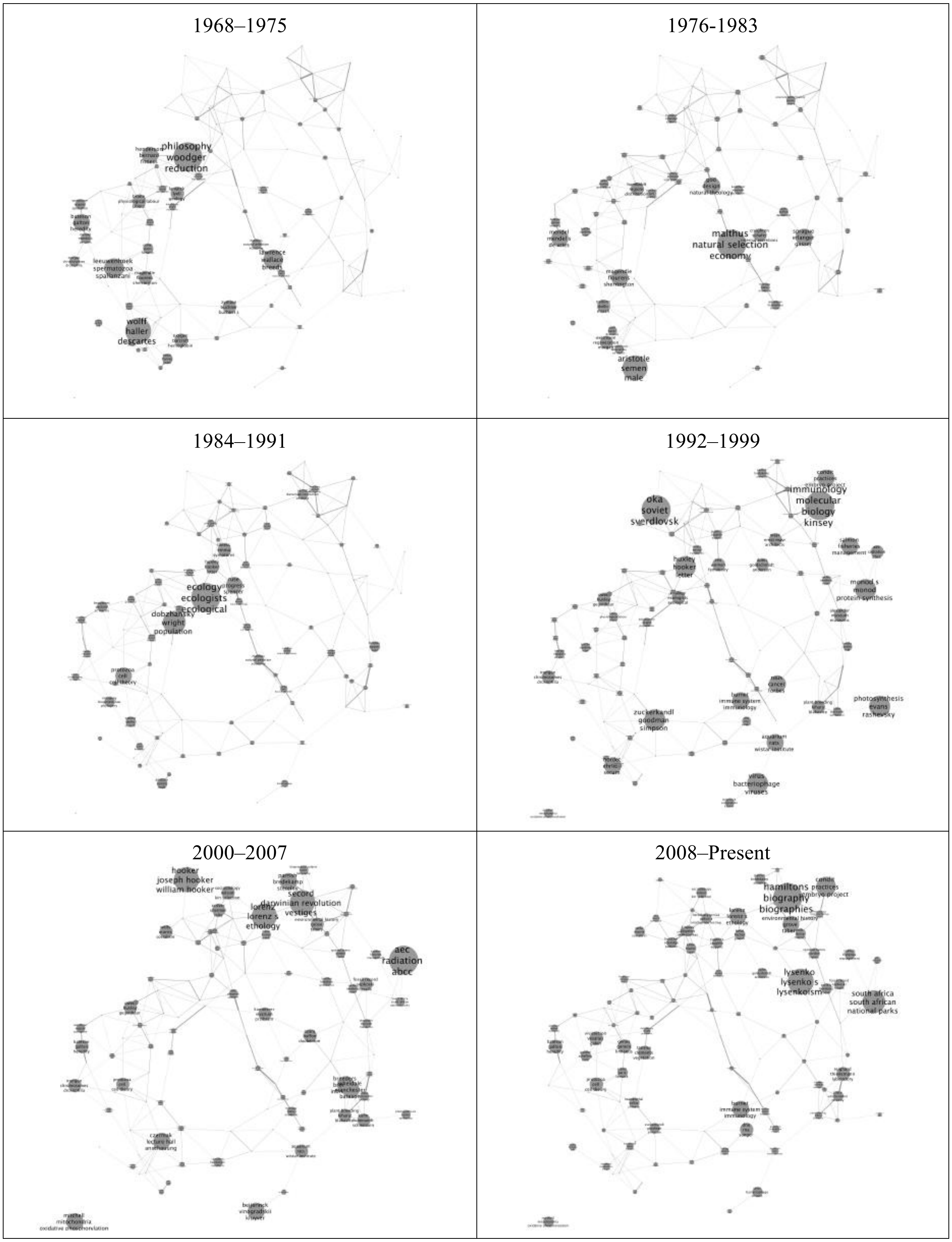

Figure 10. Graph representation of *JHB* field model. Nodes (circles) represent fields, and edges (lines) indicate that the minimum distance between fields in the t-SNE space fell below the 4th percentile; darker edges indicate closer proximity. Node size indicates the relative prevalence of each field in a given period.

In order to better visualize the temporal shifts in field representation across the entire corpus, we generated a graphical representation of the field model that reflects the thematic proximity of fields. We drew a node (circle) for each field, and drew an edge (line) between any two fields whose minimum distance in the t-SNE thematic embedding (i.e. the distance between the two closest documents across each pair of clusters) fell below the 4th percentile, with edge weights equal to the log of the inverse of distance. We calculated an optimal layout for the graph using an edge-weighted spring-embedded layout. We then generated a separate visualization of the graph for each of the six temporal divisions of the corpus, scaling node size and labels by the relative representation of each respective field in that period. The resulting series of graph visualizations are shown in figure 10. We note two significant observations when examining those visualizations in series: First, it appears that increases or decreases in representation between periods are not distributed randomly across the graph; most noticeably, there appears to be an overall shift in density from the lower left quadrant of the graph toward the upper right quadrant of the graph. When we compare the latest period (2008–2015) to the earliest period (1968–1975), we find that if the representation of a field increased overall its neighboring fields were about 2.5 times more likely to have increased. This suggests that fields are not increasing or decreasing in prominence independently over time, but that there may be a comparatively smaller number of processes that are giving rise to the observed thematic changes. Second, although density is concentrated in the upper right quadrant in later periods, there are also a greater number of fields with low but non-zero representation compared to earlier periods, suggesting a possibly higher level of thematic diversity.

### 3.3.4. Thematic diversity

In earlier sections, we evaluated whether or not the diversity of *JHB* has changed over time with respect to both geography (of content or of authors) and organisms. Similarly, the directional shifts that we observe in the representation of specific fields within *JHB* prompt us to ask whether or not the overall representation of fields is becoming more or less diverse over time. As in the case of biodiversity, the classes (fields) of interest are not equally distinct from each other. That is, some fields are more closely related than others: for example, we would not interpret an issue of *JHB* containing papers about Darwin, Lamarck, adaptationism, and evolutionary ecology to be nearly so diverse as an issue containing papers about Darwinism, aquaria, epidemiology, and forest management. Since similarities among fields are already well-described in the weighted graph representation described above, we applied the same diversity metric used to estimate taxonomic diversity (eq. 4) to estimate field diversity in each of the six eight-year periods.

Diversity in the representation of thematic fields (figure 11) was relatively flat across the first three periods (1968–1991). During the fourth and fifth periods (1992–2007) diversity was slightly higher than before, although the large bootstrap intervals make it difficult to say so definitively. The most recent period (2008–2016), however, had a significantly higher level of diversity than the earliest three periods, and also likely higher diversity than during the fourth and fifth periods. In other words, *JHB* has become more diverse over time with respect to the system of fields individuated in this model, and most of the increase in diversity occurred in recent years.

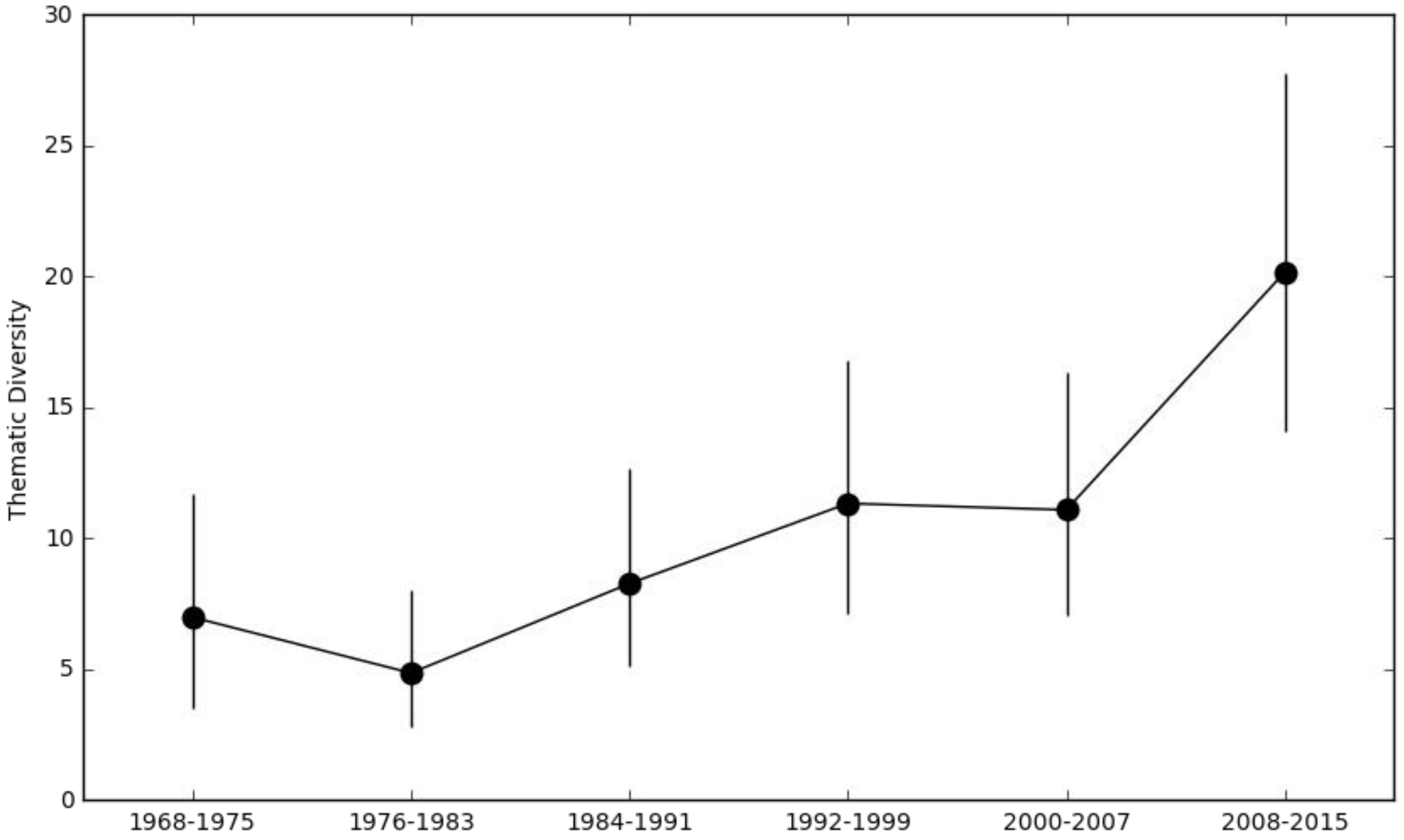

Figure 11. Thematic diversity in *JHB* over time. Thematic diversity was calculated as in equation 4, using the weighted field graph described in section 3.2.3. Higher thematic diversity indicates a greater number (richness) of fields present in a period with higher average thematic distinctness. Articles published in *JHB* became more thematically diverse over time, with the bulk of that increase occurring in the most recent period (2008–2015).

## Summary

Scholars can use probabilistic topic modeling as a mechanism for making meaningful thematic comparisons within large collections of documents. We leveraged the distribution of topics inferred from the topic model to identify clusters of thematically similar articles published in *JHB*. Our analysis suggests that the relative representation of fields within *JHB* have varied over time in ways that are consistent with directional changes in the thematic orientation, either of the discipline of history of biology as a whole or of the preferences of the journal's editors. A preliminary analysis suggests that thematically similar fields were more likely to change in similar ways over time; this could be explained by positing one or several processes acting on fields in similar ways, such as a broad change in historiographical orientation. It may be informative to incorporate additional metadata, such as citations and author characteristics, to further clarify the kinds of changes taking place within and among fields. Further research should focus on developing more sophisticated topic models that incorporate different historiographic perspectives.

The foregoing discussion showed some ways that we can begin to use tools like topic modeling to pose questions about high-level patterns in collections of documents. It is important to bear in mind, however, that nothing about the design of the various models and procedures used here are specific to the history of science (nor to history, nor to science). The most interesting and potentially transformative work in this field therefore is yet to come, as historians and philosophers of science increasingly engage in text-based modeling and other forms of quantitative analysis.

# 4. Web Application

In conjunction with the publication of this article, we will launch an interactive web application—*JHB Explorer* (https://jhbexplorer.org)—driven by the data presented here as well as several additional data sources in the growing ecosystem of digital history and philosophy of science.[18] In this section we briefly describe some of the main features of the application. We will continue to refine and expand the application as new data become available, and in response to user feedback. Note that the figures included in this section depict a prototype version of the application, and the first public release may differ visually in some respects.

The *JHB Explorer* application is comprised of both server- and client-side components. The server-side application was developed in the Python programming language (https://www.python.org/), using the Django web framework (https://www.djangoproject.com/). Client-side components were developed in the JavaScript programming language (http://www.ecma-international.org/publications/standards/Ecma-262.htm), with heavy use of jQuery (https://jquery.com/) and D3 (https://d3js.org/). All of the source code is available on GitHub (https://github.com/upconsulting/jhb-explorer) under the GNU GPL v2 open source license. Readers are invited to contribute to the project by submitting bug reports and feature requests in the GitHub issue tracker (https://github.com/upconsulting/jhb-explorer/issues). Those who wish to contribute new features or improve existing components are also very welcome to contribute code by forking the repository and submitting a pull request.

The application is roughly divided into three main parts, each corresponding to the three preceding sections in this article.

## 4.1. Geography.

We used the MapBox API (https://www.mapbox.com/) to overlay our geographic annotations on a 2D global base map. A corpus-level view shows the distribution of articles in space, and allows the user to limit the display based on time period. Clicking on a location reveals the specific articles in the selected period that were tagged with that location. We have included a feature that allows users to update our geographic annotations of individual articles, either by contributing new or more specific annotations, or by flagging our annotations as problematic. Geographic components are also included in other parts of the application, e.g. in document and author detail views (below).

## 4.2. Organisms.

We draw on linked data from the NCBI Taxonomy database to provide several views onto organisms in *JHB* (figure 12). The primary interface shows the relative distribution of taxa among the main divisions (see section 2) in the Taxonomy database for a user-selected time window as a pie-chart. Selecting one of those divisions filters a list of taxa on the right-hand side of the page that shows the number of mentions of that taxon in the current period. When a user clicks on a name in that list, they are taken to a detail view for that taxon.

---

[18] The *JHB Explorer* application was inspired by the Signs@40 online exhibit (http://signsat40.signsjournal.org) developed for the journal Signs, but incorporates a broader dataset and a larger selection of features.

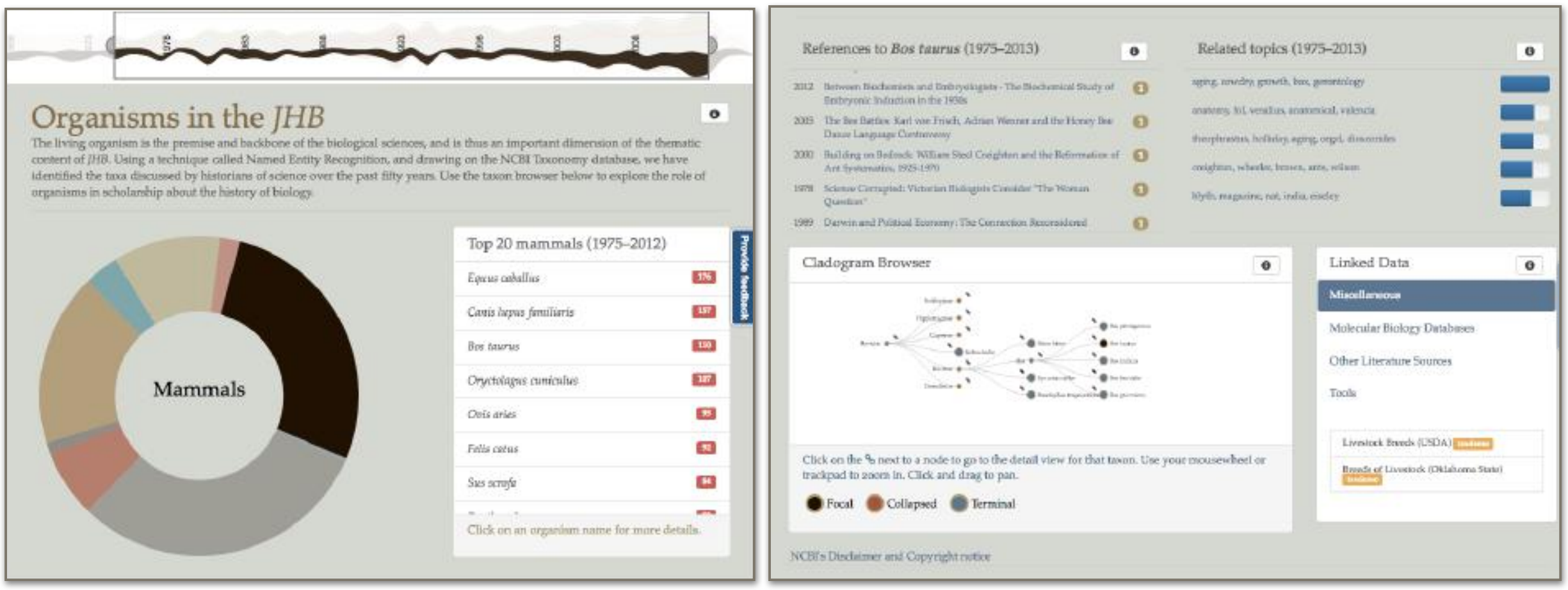

Figure 12. Organism views in *JHB Explorer*. In the summary view (left), users can examine the temporal distribution of references to taxa in each of the NCBI Taxonomy database's major divisions, as well as the specific taxa in each division. In the taxon detail view, users can see which documents refer to a taxon, associated topics, and linked data about that taxon from NCBI. The focal visualization of the taxon detail view is a cladogram browser, in which users can view related taxa.

The taxon detail view combines data from the LINNAEUS organism annotations in *JHB* articles with linked data from the NCBI Taxonomy database. A link to the NCBI Taxonomy detail view is provided in the upper-right hand corner of each page. Specific articles mentioning that taxon in the current period are shown, linking to the detail view for respective articles. There are also reciprocal links from the individual article detail view back to the detail views for taxa mentioned in that article. Topics (see section 3) with which mentions of that taxon are most strongly associated are also displayed. Linked data about the organism—such as literature sources, gene sequences, and information about voucher specimens—are displayed with links to the corresponding external record. The focal visualization of the taxon detail view is a "cladogram browser" that shows the current organism in its proximate taxonomic context. Users can navigate the cladogram by expanding and collapsing taxon nodes, and link directly to the detail view for each taxon. Note that detail views exist for taxa at all ranks: for ranks above the species, data from endotaxa are aggregated to generate a summary view (e.g. to see all mentions of Streptophytes in *JHB*).

## 4.3. Topic model.

We used the topic models and clustering techniques described in section 3 to generate two families of visualizations of the *JHB* corpus as a whole. The user may select to use any of the six topic models that we fit for this project, each of which results in a different set of visual representations.

We created a topic-centric view (figure 13, top) using a graphical representation of each topic model. In those views, each topic is represented as a node, and an edge between two nodes indicates that the two topics occur together on individual pages much more likely than we would expect by chance alone.[19] Clicking on a node reveals details about the topic, including links to documents in which the topic occurs.

[19] Specifically, we use pointwise mutual information, which relates the observed probability of two topics occurring together to the expected probability of their co-occurrence based on their independent probabilities of occurrence.

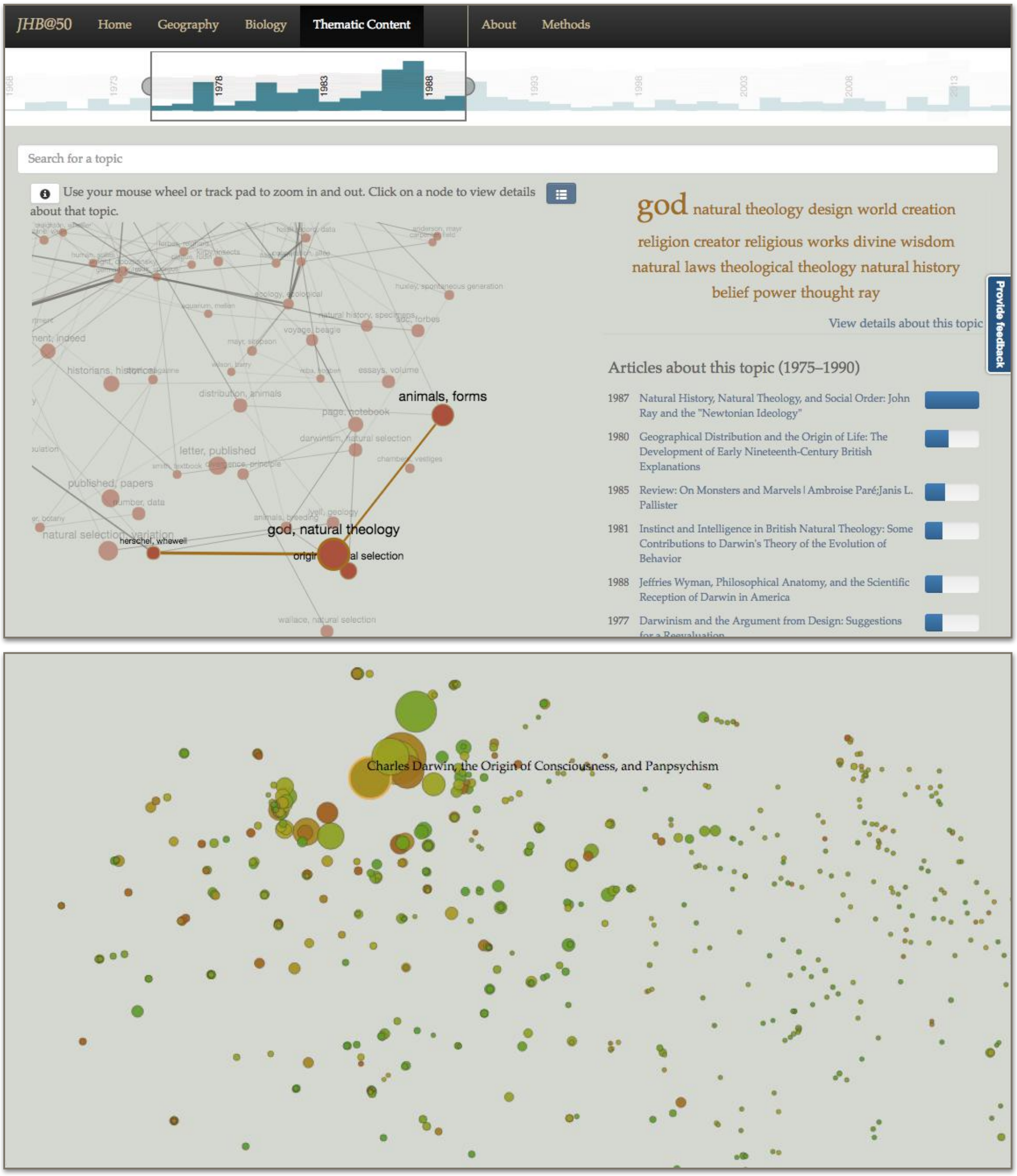

Figure 13. Two topic-driven views in *JHB Explorer*. In the top panel, topics are shown as nodes in a graph model, in which an edge between two topics indicate an improbably high probability of co-occurrence on the same pages. Users can limit the view using the time-window slider at the top of the screen, which also shows the representation of the selected topic over time. Users can also select a document-centric visualization (bottom panel), which shows a t-SNE projection of documents in topic space, with similar documents placed closer together.

We generate a separate graph representation for each potential temporal selection (i.e. for each possible start and end date), resulting in 1,225 graphs for each topic model. In order to provide a comprehensible visual experience, we generated layouts for those graphs using a linked force directed algorithm. This extends the classic force-directed layout (Fruchterman and Rheingold 1991) by incorporating attractive forces from identical nodes in temporally adjacent graphs. The resulting node positions thus aspire toward the optimal layout for each graph in isolation while simultaneously minimizing the distance that each node moves between time-indexed graphs.

We also created a document-centric view, using the t-SNE embeddings described in section 3.3.3. The main visual component (figure 13, bottom) depicts each document as a node. Documents that contain the same topics tend to be located more closely together. The user can limit the displayed documents by publication date. Clicking on a node reveals details about that document, including a link to its detail view and to topic detail views. Figure 14 shows an example of the topic detail view, in which the user can view other documents containing the topic, authors whose work contains the topic, and other related topics.

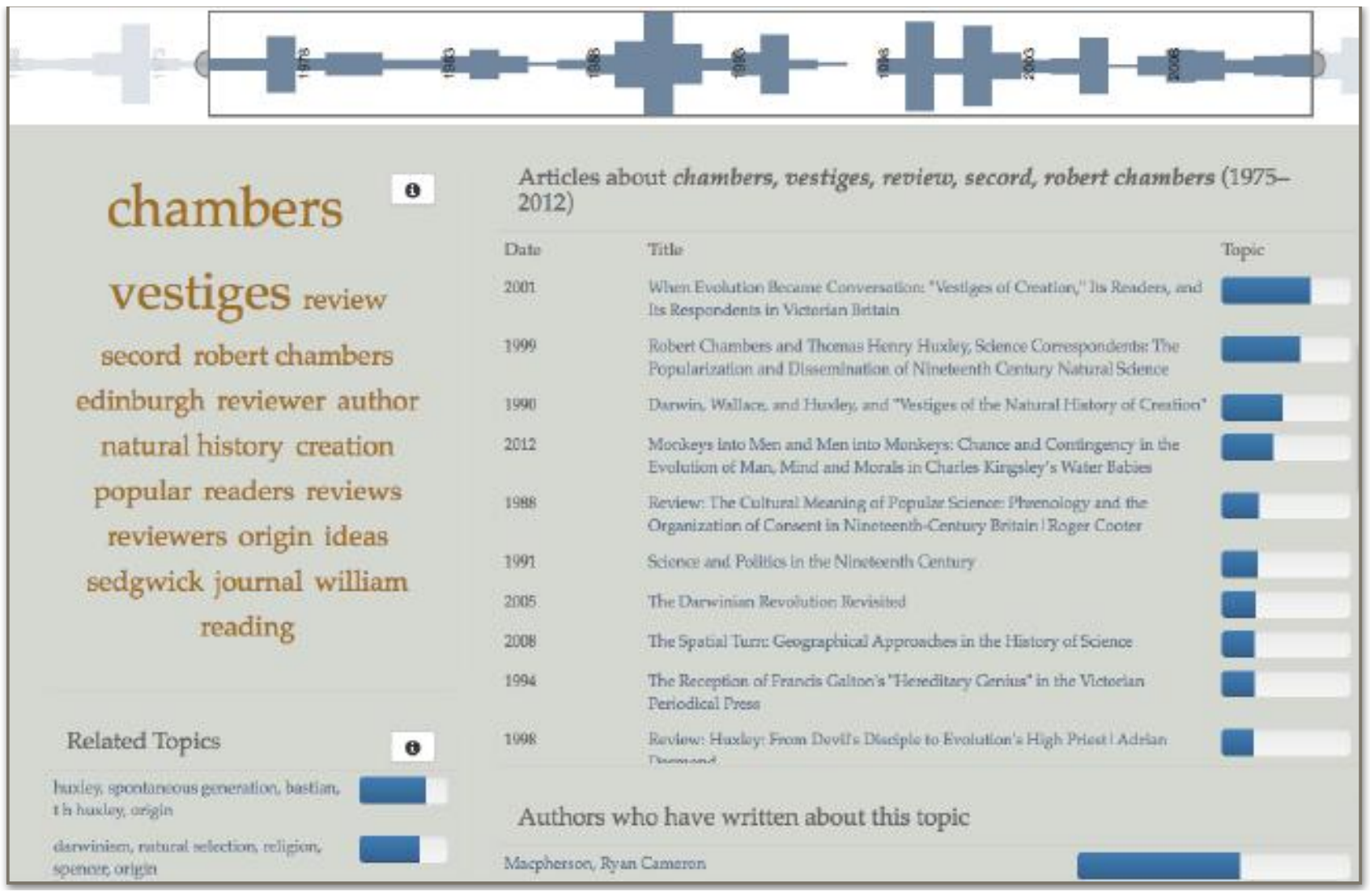

Figure 14. Topic detail view in *JHB Explorer*. The distribution of the topic over time is shown at the top of the screen. The view displays articles in which the topic occurs, as well as related topics (based on co-occurrence), and related authors.

## 4.4. Other Views

In the document detail view (figure 15), users can see how each topic in a document is distributed over that document's pages. This view also shows any detected references to organisms, and any citations to or from other articles in *JHB*. A link to a digital version of the article in either the JSTOR or Springer database is provided on each page. The author detail view (figure 16) shows articles written by a particular author, the topics that they tend to write about most, and other authors who write about similar topics. Each page includes a link out to the IsisCB Explore authority record for that author, which includes additional publication information (beyond *JHB*).

## 4.5. Future Development

We plan to develop *JHB Explorer* beyond the first public release by incorporating additional features that leverage data within and about *JHB* articles. The VogonWeb text annotation platform contains structured data about several thousand historical events relationships described in *JHB* articles, largely focused on interpersonal relationships (collaborations, friendships, correspondences) and institutional affiliations (university attendance, employment, etc.). Each of these records includes the positions in each article that substantiate or describe the relationship or event, temporal context, metadata about the contributing annotators, and references to specific authority records in the digital HPS ecosystem (*IsisCB Explore*, Virtual Internet Authority File, Conceptpower, etc.). Our goal is to incorporate summary information about those data as a continuously updated feed on each document detail view, as well as a new corpus-level view featuring a filterable graph visualization. Incorporating computer-aided human annotations of article content can provide new ways of finding relevant information in the secondary literature.

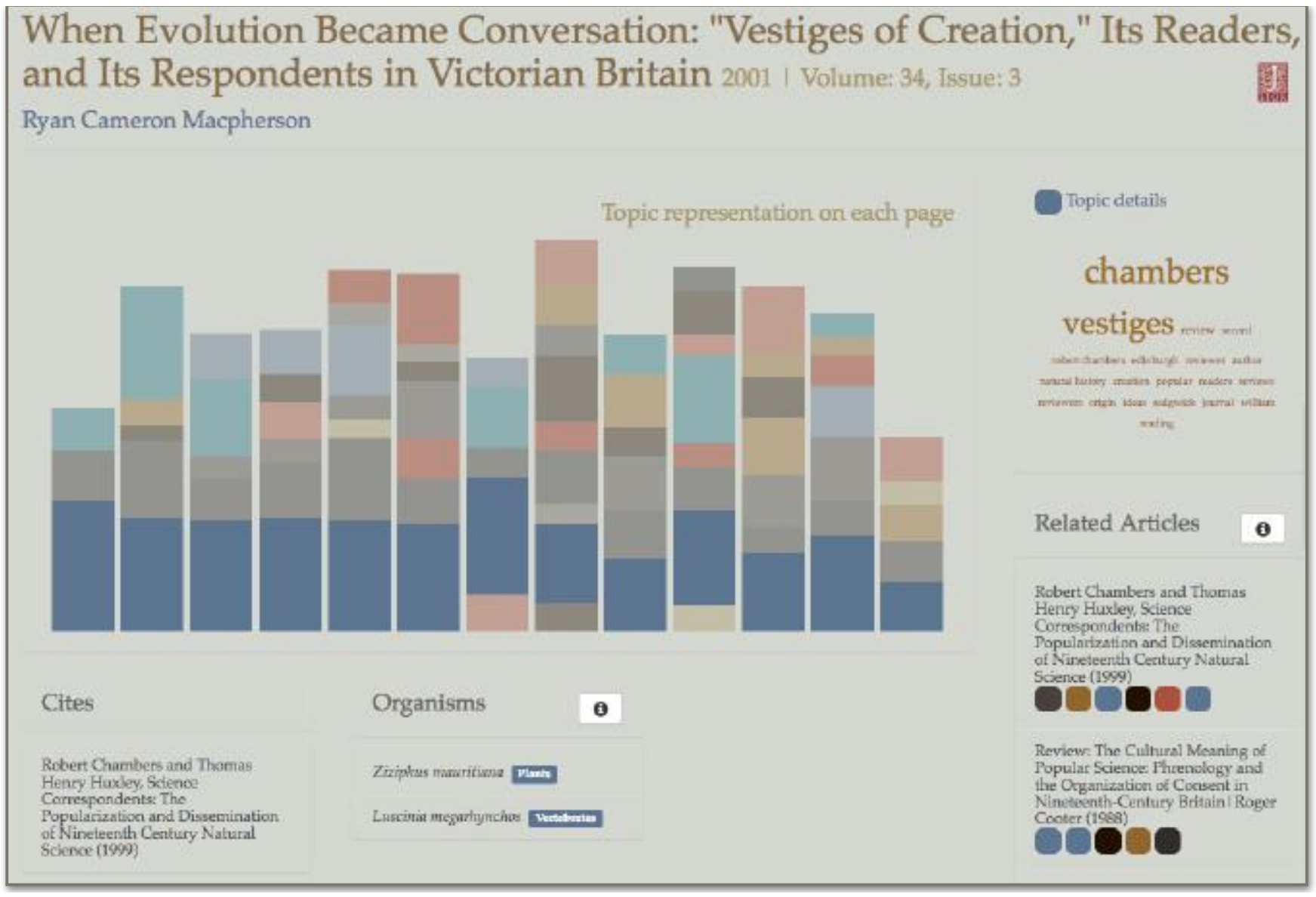

Figure 15. Document detail view in *JHB Explorer*. The focal visualization is a bar-chart showing the distribution of topics over individual pages in the article. Users can select regions of the bar chart to view details about each topic. Cited (or citing) articles are also displayed, along with any organism references. Each document view links to a digital version of the document in JSTOR or the Springer database.

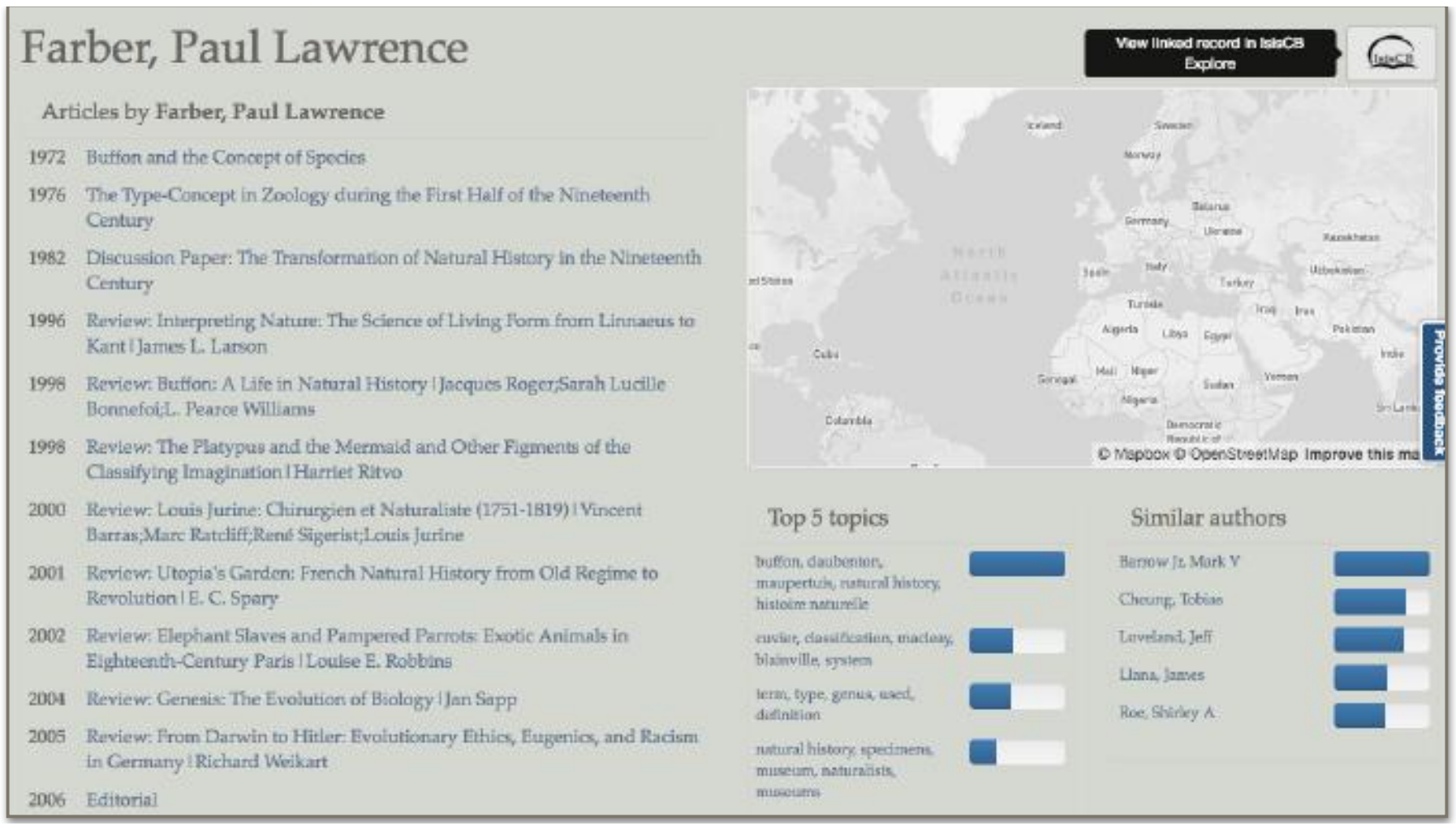

Figure 16. Author detail view in *JHB Explorer*. Users can view articles written by a particular author, the topics that occur in their articles, and similar authors (based on topic distribution).

## 5. Discussion

Quantitative methods based on a new "turn towards (big) data" are an increasingly important part of the scholarly toolkit across the humanities, and can provide new insights into patterns and processes in large collections of digitized texts. In this paper, we have employed several relatively simple techniques for extracting features of interest across the fifty-year journal run of *JHB*, ranging from manual annotation of articles with geographic information to more abstract dimensionality-reduction techniques applied to distributions of words and phrases. We then used those feature distributions to calculate metrics that describe the geographic, taxonomic, and thematic tendencies of the content of *JHB* over time. The picture that emerges from those preliminary analyses is complex. While *JHB* is not necessarily a representative sample of the history of biology (or history of science) literature more broadly, we suggest several hypotheses about the history of biology based on the analysis of JHB as a sample that warrant further quantitative and qualitative investigation:

1. The history of biology is overwhelmingly occidental in its geographic orientation, and gains in the geographic diversity of contributing authors have not yet generated a commensurate geographic diversification of content. In other words, there is a disciplinary undercurrent that maintains a western (and, frankly, northern) geographic orientation despite participation from historians in a broadening array of countries. This may not be terribly surprising to some observers of the field but, if true, the implications should be taken seriously as we consider future directions for research.
2. Our discipline may have (over)reacted to the broader attenuation of diversification of experimental organisms in the life sciences. There is some quantitative evidence of a plateau, but not a decline, in the taxonomic diversity of experimental organisms in the biomedical research literature starting around the late 1990s. The severe decline in taxonomic diversity that we observe in *JHB* around that time, however, indicates that historians may have attended quite closely to that shift and consequently focused their attention on a taxonomically narrower set of research programs. While a link, if there is one, between research trends in biology and the contemporary prioritization of research in the history of biology may or may not be desirable, it would nevertheless reinforce the need for critical reflection on and contextualization of our collective scholarly activities.
3. The history of biology may be undergoing, or has recently undergone, a significant expansion in topical scope. When we consider the content of *JHB* from the perspective of regularities in the distribution of words and phrases, we find not only that thematic tendencies rise and fall (as we would expect of any scholarly field), but that the overall diversity of those thematic tendencies was strikingly higher in recent years. This could be explained by a broadening conception of our scholarly domain, e.g. including research explicitly related to agriculture and conservation.

Further research is needed in each of these areas that leverages broader datasets and comparisons, more focused quantitative work, and linked qualitative analysis. In particular, future research should place the trends observed within *JHB* in the context of other journals related to the history of biology and history of science, technology, and medicine more broadly. Digital resources and platforms such as *IsisCB Explore* (https://data.isiscb.org) will make this increasingly practical.

Contemporary historians of science understand the promise of computation in scholarly research. Or, perhaps more accurately, they understand that there *is* promise, but perhaps not precisely *what* that promise might be. Both at professional meetings and in private conversation we are frequently asked what unique and surprising results can be generated or what new questions asked using computational methods. Those questions miss the true significance of the computational turn in the humanities: as scholars working in the 21st century we are already using computation to enhance and expedite our research. Whether aware of it or not, the vast majority of historians of science use all or nearly all of the technologies described in this paper directly or indirectly on a regular basis as they rely on various web-tools for their research (have you used Google recently?). Moreover, those questions inappropriately ascribe intrinsic epistemic value to computational methods themselves. Indeed, there are a plethora of excellent models and algorithms available to the historian to deploy on scholarly texts and datasets; where there are shortcomings it is almost certainly a failure of imagination, for we generally lack sophisticated theories that connect our disciplinary concepts to mathematical abstractions.

A more pressing and truly more interesting question is the extent to which historians of science will drive the production of tools, and the generation of computationally tractable theories, that advance our scholarly activities and that respect our disciplinary modalities of research and expression. As we have argued elsewhere (Peirson *et al.* 2016), the future of digital history of science rests not on boutique apps for text mining and visualization but rather on a strong collaborative relationship with applied mathematics and software engineering that can generate relevant and appropriate computational tools. We hope that this article will give the practicing historian of science a useful introduction to some technologies that their peers are attempting to sculpt—in collaboration with software developers—into useful and appropriate resources for the history of science. We intend for this paper to serve as an invitation to other historians of biology to participate in developing computational tools that can provide new perspectives on the development of the life sciences and on the rich corpus of scholarly literature that we ourselves have collectively generated over the past half-century.

As a final thought we would like, as historians of science, to draw a comparison between the development of our field—the history of biology—and trends within the biological sciences themselves. The availability of large datasets in the wake of the human genome project and the subsequent -omics revolutions triggered the emergence of big data driven computational methods throughout the life sciences. Some overly enthusiastic practitioners and commentators have heralded those developments as the beginning of new era, where sophisticated analyses of patterns within large datasets replace painstaking (and slow) causal and experimental analysis. Needless to say, this exuberance turned out to be irrational. What became clear, however, is that there is a fruitful symbiosis between data-driven computational approaches and detailed experimental studies. We see a similar complementarity between the patterns revealed by computational studies in the history of science and the very sophisticated (and time consuming) work that has characterized our field. As in biology, being able to embed such detailed studies within broad patterns and comparative analyses only adds to their significance while we also get a much better understanding of the whole landscape of the life sciences and its various contexts through computational studies. With these options in mind, we are looking forward to the next fifty years of *JHB*.